\documentclass[12pt]{article}

\usepackage{a4wide}
\usepackage{xspace}
\usepackage{amsmath}
\usepackage{amssymb}
\usepackage{url}
\usepackage{hyperref}
\usepackage{graphicx}

\newcommand{\ie}{{\it i.e.}\xspace}
\newcommand{\eg}{{\it e.g.}\xspace}

\newcommand{\Rbest}{\ensuremath{R_{\rm best}}\xspace}
\newcommand{\Qa}[1]{\ensuremath{Q_{f=#1}^{w}}\xspace}
\newcommand{\dd}{\ensuremath{{\rm d}}}
\newcommand{\dm}{\ensuremath{\delta m}\xspace}
\newcommand{\dmax}{\ensuremath{\delta m_{\rm max}}\xspace}

\newcommand{\avg}[1]{\left\langle #1 \right\rangle}

\title{\textsf{Optimal jet radius in kinematic dijet reconstruction}}
\author{%
  Gregory~Soyez\\[5pt]
  \it\small CERN, Department of Physics, Theory Unit, CH-1211 Geneva 23, Switzerland,\\
  \it\small Institut de Physique Th\'eorique, CEA Saclay, CNRS URA 2306,\\
  \it\small F-91191 Gif-sur-Yvette, France\\[5pt]
}

\date{}

\begin{document}

\maketitle

\begin{abstract}
  Obtaining a good momentum reconstruction of a jet is a compromise
  between taking it large enough to catch the perturbative final-state
  radiation and small enough to avoid too much contamination from the
  underlying event and initial-state radiation. In this paper, we
  compute analytically the optimal jet radius for dijet
  reconstructions and study its scale dependence. We also compare our
  results with previous Monte-Carlo studies.
\end{abstract}

\newpage

\tableofcontents

%%%%%%%%%%%%%%%%%%%%%%%%%%%%%%%%%%%%%%%%%%%%%%%%%%%%%%%%%%%%%%%%%%%%%%%%%%
%% introduction
%%%%%%%%%%%%%%%%%%%%%%%%%%%%%%%%%%%%%%%%%%%%%%%%%%%%%%%%%%%%%%%%%%%%%%%%%%
\section{Introduction} \label{sec:intro}

With the start of the LHC as a main source of motivation, jet physics has seen a
tremendous development over the last couple of years. A significant
amount of effort has been placed in trying to define jets in an
optimal way for new physics searches. 
Just to mention a few examples, this includes the introduction of new
jet algorithms \cite{siscone, antikt}, new techniques to clean the
Underlying Event (UE) contribution to the jets \cite{boosted_higgs,
  optimisation, pruning, trimming}, or a large series of sub-jet techniques
aimed at tagging boosted objects
\cite{boosted_higgs,pruning,boosted_ww,boosted_top,boosted_top_tw,boosted_chi,boosted_tth,boosted_susyh}.

An aspect that has been slightly less investigated is the values that
one should use for the parameters inherent to a jet definition. A
noticeable example is the case of the ``radius'' of a jet, often
denoted by $R$, common to all the standard jet definitions used in
hadronic collisions. 
It has been noticed \cite{optimisation} that if one wants to optimise
the kinematic reconstruction of dijet events at the LHC, it is
mandatory to adapt the radius of the jet when varying the scale of the
process and the type of partons involved in the reconstruction.

With the appearance of new techniques, new parameters are added and
one might expect the corresponding multi-dimensional optimisation to
become more and more involved.
There is therefore a risk that a poor choice of parameters for the
definition of jets counteracts the positive effect of these new
refinements.

In this paper we thus want to take an alternative, complementary,
approach and, instead of adding a new procedure with its own
parameters, study how it is possible to fix one parameter in an
optimal way given the properties of the events we want to reconstruct.
In other words, we shall remove one free parameter by determining it
analytically from the scales in the process we shall look at.
Practically, we shall focus on the determination of the most natural
parameter, the radius $R$, for the reconstruction of a massive colour-neutral
object decaying into two jets, as done within a Monte Carlo approach
in \cite{optimisation}. In that case, we want to choose $R$ large
enough to catch the perturbative QCD radiation, but not too large to
avoid an excessive contamination from the UE. Our approach thus goes
along the same line as \cite{analytic}, except that we shall extend
the discussion to include extra features requested by the fact that we
are looking at a peaked distribution and wish to obtain a more precise
determination of $R$.
Contrarily to the case of \cite{VR}, where an attempt was made to
adapt dynamically the size of the jet with its hardness by replacing
the size parameter with a dimensionful scale varying with the process
under consideration, we shall determine a unique value for $R$ from
the hard scale of the event and the scale of the UE, \ie not
introducing any alternative parameter\footnote{In the case of
  \cite{VR}, this would mean predicting the scale parameter as a
  function of known properties of the events.}.

The paper is organised as follow: in Section \ref{sec:process} we
describe the process we will study as well as the details on how we
proceed with the event analysis. We also describe in that Section our
strategy for the analytic computation of the optimal radius \Rbest.
We then proceed with the computation of \Rbest itself. Section
\ref{sec:pqcd} concentrates on the situation at the partonic level,
where only perturbative radiation has to be taken into account; while
in Section \ref{sec:UE} we add to the picture the contamination due to
the UE. In both cases, we shall compare our results with what we
obtain from Monte-Carlo studies, as in \cite{optimisation}. Finally,
in Section \ref{sec:subtraction}, we shall discuss the extraction of
the optimal jet radius for situations where we perform an underlying
event background subtraction.

%%%%%%%%%%%%%%%%%%%%%%%%%%%%%%%%%%%%%%%%%%%%%%%%%%%%%%%%%%%%%%%%%%%%%%%%%%
%% the process we'll study: dijet reconstruction
%%%%%%%%%%%%%%%%%%%%%%%%%%%%%%%%%%%%%%%%%%%%%%%%%%%%%%%%%%%%%%%%%%%%%%%%%%
\section{Dijet decay of a massive object} \label{sec:process}

As in \cite{optimisation}, we shall study the hard processes $q\bar
q\to Z'\to q \bar q$ and $gg\to H\to gg$, where the (fictitious) $Z'$
and $H$ are made very narrow. The main advantage of such simple
processes is that one can easily study the definition of jets at a
given scale, by varying the mass $M$ of the colourless resonance. It
also allows one to study the differences between quark and gluon jets,
and to test the validity of our computations with Monte-Carlo
simulations.

The procedure for the event analysis is straightforward: we cluster
the events with a given jet definition, select the two highest-$p_t$
jets and reconstruct the heavy object from them. In Monte-Carlo
studies, the events were generated either with Pythia \cite{pythia}
(tune DWT) or with Herwig\footnote{In the case of Herwig, we have
  actually generated $q\bar q\to {\rm Graviton} \to q \bar q$ and
  $gg\to {\rm Graviton} \to gg$ events, which does not make any
  difference in the discussions throughout this paper.}  \cite{herwig}
with the default Jimmy tune for the UE. We have required that the jets
satisfy $p_t\ge 10$ GeV and $|y|\le 5$, and we have further imposed
that the two hardest jets, the ones used to reconstruct the heavy
object, are close enough in rapidity, $|\Delta y|\le 1$ (so that the
transverse momentum of the jets remains close to $M/2$).

The quantification of the performance of a given jet definition is
also borrowed from \cite{optimisation}: we define $\Qa{z}$ as the
width of the smallest mass window that contains a fraction $f=z$ of
the reconstructed objects\footnote{In \cite{optimisation}, \Qa{z} was
  defined as a fraction of the generated objects rather than a
  fraction of the reconstructed ones. Though the former is practically
  more reliable, choosing the later does not affect \Rbest in practice
  and simplifies considerably the analytic computation.}. With that
definition, a smaller \Qa{z} would correspond to a narrower peak and
thus a better reconstruction quality. Finding the optimal $R$, \Rbest,
is therefore equivalent to finding the minimum of \Qa{z}, seen as a
function of $R$.

Our task in this paper is to perform an analytic computation of the
spectrum of the reconstructed mass peak $\dd P/\dd m_{\rm rec}$,
or $\dd P/\dd \dm$ with $\dm = m_{\rm rec}-M$, the difference
between the reconstructed and the nominal mass. By integrating that
spectrum, we can compute the probability
$P(\dm_1,\dm_2)$ for the difference \dm to be between $\dm_1$ and
$\dm_2$. If we write the quality measure as $\Qa{z}=q_2-q_1$, where
$q_1$ and $q_2$ are the lower and upper end of the mass window defining
\Qa{z}, it is a straightforward exercise to show that it can be
computed from the analytic spectrum by solving
\begin{equation}\label{eq:Qconstraints}
\left.\frac{\dd P}{\dd \dm}\right|_{\dm=q_1} =
\left.\frac{\dd P}{\dd \dm}\right|_{\dm=q_2}
\qquad\text{and}\qquad
P(q_1, q_2) = z.
\end{equation}
Repeating this for different values of $R$, one can then compute
$\Rbest$, the value of $R$ for which the quality measure is
minimal. We shall use $z=0.25$.

As far as the analytic computation itself is concerned, there are
three physical contributions that will affect the width of the
reconstructed peak: final-state radiation, initial-state
radiation\footnote{Since the produced massive object is colourless,
  there is no interferences between gluons emitted from the
  initial-state and final-state partons.} and the UE. The first two
are perturbative; while final-state radiation leads to losses by
emissions out of the jet, initial-state radiation adds extra,
unwanted, radiation inside the jet. Already at the purely partonic
level, there is thus an optimal radius. At this stage, the mass
spectrum will depend on the mass of the reconstructed object, $M$, and
whether it decays into quarks or gluons since, perturbatively, gluons
radiate more than quarks.

The third contamination, the underlying event, is softer. Like the
initial-state radiation, it tends to move the reconstructed mass
towards larger values by clustering soft particles into the jet. The
amount of contamination involves another scale: the density of UE (per
unit area) $\rho$. The most important source of dispersion in the mass
spectrum comes from the fact that $\rho$ varies from an event to
another.

In the next two Sections, we first concentrate on the purely
perturbative behaviour, then include the effect of the UE. But before
getting our hands dirty, there is one additional point that needs to
be discussed: the fact that we should also expect some effect due to
hadronisation. Actually, for the purposes of this paper, hadronisation
plays a limited role. In theory \cite{analytic}, it leads to a
loss that increases like $1/R$ at small $R$, to be compared with a
$\log(1/R)$ for the perturbative component, and so should dominate the
small-$R$ behaviour. However, the normalisation is such that for
practical values of $R$ the perturbative contribution
dominates, and the effect of hadronisation on the value of \Rbest
remains negligible.

A final comment concerns the jet algorithm that we shall consider: we
will mostly focus on the anti-$k_t$ algorithm \cite{antikt} as its
behaviour is simpler. In Section \ref{sec:full_algs} we shall discuss
the case of the Cambridge/Aachen \cite{cam} and Cambridge/Aachen with
filtering \cite{boosted_higgs,optimisation} algorithms.

%%%%%%%%%%%%%%%%%%%%%%%%%%%%%%%%%%%%%%%%%%%%%%%%%%%%%%%%%%%%%%%%%%%%%%%%%%
%% Perturbative QCD spectrum
%%%%%%%%%%%%%%%%%%%%%%%%%%%%%%%%%%%%%%%%%%%%%%%%%%%%%%%%%%%%%%%%%%%%%%%%%%
\section{Perturbative QCD spectrum} \label{sec:pqcd}

The part of the spectrum that is the most naturally computed is the
pure perturbative QCD contribution. This is equivalent to concentrating
on partonic events. In that case, two types of contribution can
affect the reconstruction of the jet momentum: final-state radiation
outside of the jet which leads to an underestimation of the jet
momentum, and in-jet initial-state radiation yielding an
over-estimation of the momentum.

In this Section, we shall thus compute the pure partonic spectrum. We
do this at the one-gluon-emission (OGE) level. Since we are interested
in the behaviour of the spectrum close to the mass peak, we are
dominated by the $1/z$ singularity directly coming from the soft
divergence of QCD. In that region, to get an integrable spectrum
around the mass peak, it is also important to resum the
$(\alpha_s \log(1/z))^n$ contributions to all orders, which we shall
do assuming a Sudakov-like exponentiation, neglecting non-global
logarithms \cite{nonglobal,mathieu,ngjetdef}.

For the clarity of the computation, we first deal with the soft
approximation, where we only keep the $1/z$ part of the gluon
emission. We then introduce the sub-leading corrections coming from the PDF
in the initial state which can have non-negligible effects, especially
in the case of gluon resonances at large mass.

% Sotf-gluon emission approximation
%%%%%%%%%%%%%%%%%%%%%%%%%%%%%%%%%%%%%%%%%%%%%%%%%%%%%%%%%%%%%%%%%%%%%%%%%%
\subsection{Soft-gluon emission approximation} \label{sec:base}

Because of longitudinal boost invariance, we can assume that the heavy
object is produced at $y=0$. Furthermore, because of our cut $|\Delta
y|\le 1$ on the dijets, we shall assume that the decay products are at
the same rapidity. The correction due to the finiteness
of $\Delta y$ can be computed but has very little effect (at most a
few percent). Since the calculation for a nonzero $\Delta y$ cut becomes rather technical, we shall keep things
simple and assume $\Delta y=0$. 
At the lowest order of perturbation theory, the incoming ($p_{1,2}$)
and outgoing ($k_{1,2}$) partons thus have the following momenta
\begin{align*}
  p_1^\mu & \equiv \frac{M}{2}\left( 0, 0, 1, 1 \right) \\
  p_1^\mu & \equiv \frac{M}{2}\left( 0 ,0,-1, 1 \right) \\
  k_1^\mu & \equiv \frac{M}{2}\left( 1, 0, 0, 1 \right) \\
  k_2^\mu & \equiv \frac{M}{2}\left(-1, 0, 0, 1 \right) 
\end{align*}
We shall consider the emission of an extra soft gluon:
\[
  k^\mu \equiv \frac{zM}{2} \left(\cos(\phi),\sin(\phi),\sinh(y),\cosh(y)\right)
\]
with $z\ll 1$. The probability for such an emission to happen is given
by the antenna formula
\begin{equation}\label{eq:antenna}
\frac{\dd P}{\dd^4k} = 2 C_R\frac{\alpha_s}{2\pi^2}\frac{(p_1.p_2)}{(p_1.k)(p_2.k)}
                       \delta(k^2)
\end{equation}
for an emission from the initial state, and a similar expression with
$p_i$ replaced by $k_i$ for final-state radiation. The pre-factor
$C_R$ is a colour factor which should be equal to $C_F$ for an
emission from a quark line and $C_A$ for an emission from a gluon line.

Note that in the case of a single soft emission, the gluon is caught if and
only if it is at a geometric distance smaller than $R$ from one of the
two ``leading'' partons in the final state\footnote{This would be true
for all the recombination-type algorithms, like $k_t$ \cite{kt} or
Cambridge/Aachen, with or without filtering, as well as for the
SISCone \cite{siscone} algorithm (at least in the $z\ll 1$ limit).}.
 
If we denote by $\dm$ the difference, $m_{\rm rec}-M$, between the
reconstructed mass and the nominal one, initial-state radiation will
lead to an overestimation of the jet mass with $\dm = zM/2\,{\rm
  cosh}(y)$ whenever the additional gluon is radiated inside the jet,
while final-state radiation outside the jet will lead to an
underestimation of the mass with $\dm = -zM/2\,\cosh(y)$.

Integrating over the rapidity and azimuth of the emitted
gluon and using (\ref{eq:antenna}), we find
\begin{equation}\label{eq:pqcd_barespectrum_i}
\frac{\dd P^{(0)}_i}{\dd \dm} = 2 \int_{{\rm min}(\phi^2,(\pi-\phi)^2)+y^2<R^2} \dd \phi\,\dd y\:
  \frac{\alpha_sC_R}{\pi^2} \frac{1}{\dm}
 = \frac{2\alpha_s C_R}{\pi^2}\, A_i(R)\,\frac{1}{\dm}
\end{equation}
for the initial-state radiation spectrum, and
\begin{align}
\frac{\dd P^{(0)}_f}{\dd \dm} & = 2 \int_{{\rm min}(\phi^2,(\pi-\phi)^2)+y^2>R^2} \dd\phi\,\dd y\:
  \frac{\alpha_sC_R}{\pi^2} \frac{1}{|\dm|}
  \frac{1}{\cosh^2(y)-\cos^2(\phi)}
 \nonumber\\
 & = \frac{2\alpha_s C_R}{\pi^2}\, A_f(R)\,\frac{1}{|\dm|}\label{eq:pqcd_barespectrum_f}
\end{align}
for final-state radiation, with
\begin{eqnarray}\label{eq:pqcd_geomcoefs}
A_i(R) & = & \pi R^2,\nonumber\\[-3mm]
&&\\[-3mm]
A_f(R) & = & 2 \pi \log(2/R) - \frac{\pi}{72} R^4 + {\cal O}(R^8)\nonumber
\end{eqnarray}
capturing the geometric factors and the $R$ dependence.
One recognises the typical expected behaviour: the contamination due
to initial-state radiation is proportional to the jet area, \ie to
$\pi R^2$, while the logarithmic behaviour at small $R$ of the final-state
radiation is the trace of the collinear divergence in QCD. In both
cases, the $1/\dm$ behaviour corresponds to the soft divergence of QCD.

Note that the coefficient of the final-state radiation $A_f(R)$ has
been expanded in series of $R$. The logarithmic behaviour at small $R$
corresponds to the leading collinear term as computed in
\cite{analytic} but we have kept enough terms in the expansion to have
a correct description over the whole $R$ range. In particular, the
numerator in the argument of the logarithm will play a mandatory role
to allow the best radius to go above 1, which is the case in the
purely perturbative spectrum and, more generally, for large-mass gluon
resonances.

It is also important to discuss the choice of scale $\mu$ for
$\alpha_s$. When using equation (\ref{eq:antenna}), the argument of
the coupling is, for initial and final-state radiation respectively,
\begin{eqnarray}
\mu_i^2 &\!=\!& 2 \frac{(p_1.k)(p_2.k)}{(p_1.p_2)}\,=\,\frac{z^2M^2}{4},\nonumber\\[-3mm]
&&\\[-3mm]
\mu_f^2 &\!=\!& 2 \frac{(k_1.k)(k_2.k)}{(k_1.k_2)}\,=\,\frac{z^2M^2}{4}\left[\cosh^2(y)-\cos^2(\phi)\right].\nonumber
\end{eqnarray}
Assuming that emissions close to the edge of the jet are dominant, we
will use the following simplified approximation:
\begin{equation}\label{eq:couplingscales}
\mu_i = \frac{\dm}{\cosh(R)}
\qquad\text{ and }\qquad
\mu_f = \frac{R\,\dm}{\cosh(R)}.
\end{equation}

As we are interested in the behaviour of the mass spectrum in the
vicinity of the peak, the last thing we need to consider is the
virtual corrections and the higher-order terms that diverge as
$\alpha_s^n\,\log^n(1/z)$ in that region. In principle, this would
depend on the jet algorithm under consideration and an exact
computation would involve non-global logarithms
\cite{nonglobal,mathieu,ngjetdef}. In practice, we shall adopt a simpler
approach and assume that the one-gluon-emission result
exponentiates. This is equivalent to the introduction of a Sudakov
factor
\begin{equation}\label{eq:sudakov}
{\cal S}_{i,f} = \exp\left(-\int_{|\dm|}^{|\dmax|} \dd u 
\left.\frac{\dd P^{(0)}_{i,f}}{\dd\dm}\right|_{\dm=u}\right),
\end{equation}
where the integration is performed up to $\dmax=M/2$, corresponding to
$z=1$ at $y=0$. This is an arbitrary choice though a different one
would only introduce corrections beyond the level of precision we are
working at.
Using equations (\ref{eq:pqcd_barespectrum_i}),
(\ref{eq:pqcd_barespectrum_f}), (\ref{eq:couplingscales}) and
(\ref{eq:sudakov}) we find
\begin{equation}\label{eq:spectrum_pQCD_RC}
  \frac{\dd P_{i,f}}{\dd\dm} = \frac{K_{i,f}}{|\dm|\,\log(M/(2\Lambda_{i,f}))}
  \left[\frac{\log(|\dm|/\Lambda_{i,f})}{\log(M/(2\Lambda_{i,f}))}\right]^{K_{i,f}-1},
\end{equation}
with
\begin{equation}\label{eq:pqcd_coefs}
K_{i,f}=\frac{C_R}{\beta_0\pi}A_{i,f}(R),
\quad
\Lambda_i = \cosh(R)\,\Lambda_{\rm QCD}
\quad\text{and}\quad
\Lambda_f = \frac{\cosh(R)}{R}\,\Lambda_{\rm QCD}.
\end{equation}

We note that as long as $K_{i,f}$ is smaller than 1, the spectrum goes to 0 when $\dm$ tends to
$\Lambda_{i,f}$. Alternatively, we could regularise the Landau pole in
the expression for $\alpha_s$ but we have noticed no significant effect
in our conclusions when doing so.

\begin{figure}
\includegraphics[angle=270,width=\textwidth]{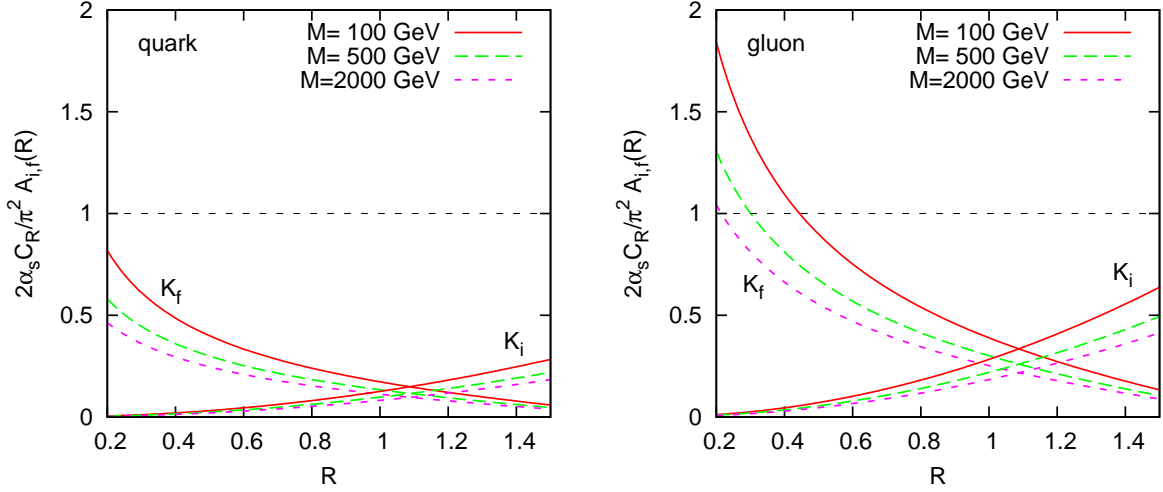}
\caption{Coefficients $2\alpha_aC_RA_{i,f}(R)/\pi^2$, governing the
  behaviour of the initial or final-state QCD spectrum as a function
  of $R$ for different masses. The left plot covers the case of quark
  jets for which we use $C_R=C_F=4/3$, and the right plot is for gluon
  jets for which $C_R=C_A=3$.}\label{fig:coefs}
\end{figure}

Together with the spectrum, we will also need the total probability
for the reconstructed mass to be within a certain range, \ie the
integrated spectrum. Denoting by $P(\dm_1,\dm_2)$ the probability
for $m_{{\rm rec}}-M$ to be between $\dm_1$ and $\dm_2$, we easily find
\begin{equation}\label{eq:proba_pqcd_rc}
P(\Lambda_{i,f},\dm) = \left[\frac{\log(|\dm|/\Lambda_{i,f})}{\log(M/(2\Lambda_{i,f}))}\right]^{K_{i,f}}
\end{equation}
corresponding to the spectrum (\ref{eq:spectrum_pQCD_RC}). Note that
in this computation, we have assume that the spectrum is cut at the
Landau pole, corresponding to $|\dm| = \Lambda_{i,f}$. 

Before proceeding with a more involved computation, it is important to
comment on the limit of validity of this perturbative
computation. From equations (\ref{eq:pqcd_barespectrum_i}), (\ref{eq:pqcd_barespectrum_f}) and
(\ref{eq:pqcd_geomcoefs}), one might expect higher-order corrections
to be important when $2\alpha_sC_RA_{i,f}(R)/\pi^2$ becomes of order
1. Fixing the scale in the coupling to $M/2$ for the sake of the
argument, we have plotted that quantity on Fig. \ref{fig:coefs} for
various cases of interest. In the case of
quark-jets, both $K_i$ and $K_f$ remain smaller than 1. 
The case of gluon jets is more interesting; we see that the
coefficient for final-state radiation goes above one at small $R$. In
the case of a small-mass resonance, this can happen for values of $R$
as large as 0.5. As a rule of thumb, we can thus say that, in the case of a
small-mass resonance decaying to gluon jets, our perturbative
computation would only be valid for $R\gtrsim 0.5$. It is particularly
interesting to relate that to the observation made in
\cite{optimisation} that in this precise case of small-mass resonances
decaying into gluons, we observed a disappearance of the peak in the
reconstructed mass spectrum for small values of $R$. Though this was
happening at slightly smaller values of $R$ (around $R\approx 0.4$),
the present computation provides an explanation of this effect: when
$2\alpha_aC_RA_{i,f}(R)/\pi^2$ grows above 1, the peak in the
final-state radiation spectrum \eqref{eq:spectrum_pQCD_RC} is no
longer around $z=0$ and the mass peak disappears. We shall come back
to this point later in Section \ref{sec:full_mc}.

Note also that, from Fig. \ref{fig:coefs}, one can have a feeling of
what the optimal radius will be for parton-level events. Indeed, one
can expect that, for $R=\Rbest$, the spectrum will be nearly symmetric
around the nominal mass (\ie around $z=0$), that is $K_i(\Rbest) =
K_f(\Rbest)$. For both quarks and gluons and for all masses, this
means that one should expect $\Rbest \approx 1.1$. We will see later
on that this is indeed relatively close to what we shall obtain but
that PDF effects, computed in the next Section, induce some departure
from that situation.

% PDF effects
%%%%%%%%%%%%%%%%%%%%%%%%%%%%%%%%%%%%%%%%%%%%%%%%%%%%%%%%%%%%%%%%%%%%%%%%%%
\subsection{PDF effects} \label{sec:pdf}

In order to get a better description of the variety of processes we
consider over the whole kinematic range, we shall see that it is
important to take into account the effect of the PDF on 
initial-state radiation. This addresses the fact that the emission of
an initial-state gluon imposes to take the parton distribution
functions at a larger momentum fraction $x$. This should therefore
introduce an additional suppression, more important at large mass and
for gluon-jets than for quark jets.

Let us consider the production of a heavy particle at a fixed rapidity
$y$. The leading-order cross section is easily obtained:
\begin{align}
\sigma_a^{LO}(y)
 & = \int {\rm d}x_1 {\rm d}x_2 \: f_a(x_1) \, f_a(x_2)\:
     \delta(x_1x_2s-M^2) \,
     \delta\left(\frac{1}{2}\log\left(\frac{x_1}{x_2}\right)\!-y\!\right)
     \hat\sigma(M,y)
      \nonumber\\
 & = \frac{1}{s}\, f_a\left(\frac{M}{\sqrt{s}}e^{ y}\right)
                   f_a\left(\frac{M}{\sqrt{s}}e^{-y}\right)\,\hat\sigma(M,y),
\end{align}
where the index $a$ represents the colour of the colliding partons,
$f_a(x)$ is the PDF implicitly taken at a scale $M$, and $s$ is the
centre-of-mass energy of the collision. This expression also comes
with the kinematic constraint $|y|\le \log(M/\sqrt{s})$.

We should now add to that process an extra parton emitted at a
rapidity $\eta$. For our purpose of computing the
initial-state radiation contamination to a jet, we can safely make the
simplifying assumption that the rapidity of the gluon is the same as
the rapidity of the ``leading'' parton, \ie, $\eta=y$. By doing so, we
decouple the effects of the PDF from the effects of the jet
clustering. The latter will be reinserted later on by multiplying our
results by the geometric factor $A_i(R)$. We thus have
\[
\sigma_a^{NLO}(y) =  \frac{\alpha_s}{2\pi^2}
  \int {\rm d}x_1 {\rm d}x_2 {\rm d}\xi \, 
  f_a(x_1) f_a(x_2)\, \frac{2 C_a}{1-\xi}\,
  \delta(x_1x_2\xi s-M^2) \,
  \delta\left(\frac{1}{2}\log\left(\frac{x_1\xi}{x_2}\right)\!-\!y\right)\hat\sigma.
\]
In that expression, $\xi$ is the longitudinal fraction of $x_1$
carried by the parton entering the collision satisfying\footnote{We only keep the
dominant soft-gluon emission in the splitting function.} $1-\xi\ll 1$. If we
rewrite it in terms of the longitudinal momentum of the emitted parton
(measured w.r.t. the momentum of the beam) $\zeta = (1-\xi) x_1$, one
gets after a bit of algebra
\begin{equation}
\frac{{\rm d}\sigma_a^{NLO}}{{\rm d}\zeta}(y) = 
  \frac{1}{s}\, \frac{\alpha_s}{2\pi^2}\,
  f_a\left(\frac{M}{\sqrt{s}}e^{ y}+\zeta\right)
  f_a\left(\frac{M}{\sqrt{s}}e^{-y}      \right)
  \frac{2 C_a}{\zeta}\hat\sigma.
\end{equation}

Keeping the kinematic conventions introduced in the previous
section, the longitudinal fraction $\zeta$ can be rewritten in terms
of the transverse momentum $k_t=zM/2\approx \dm$ of the parton as
\begin{equation}
\zeta = \frac{e^y}{\sqrt{s}}\,\delta m,
\end{equation}
which finally implies
\begin{equation}
\frac{{\rm d}\sigma_a^{NLO}}{{\rm d}\dm}(y) = 
  \frac{1}{s}\, \frac{\alpha_s}{2\pi^2}\,
  f_a\left(\frac{(M+\dm)e^{ y}}{\sqrt{s}}\right)
  f_a\left(\frac{ M     e^{-y}}{\sqrt{s}}\right)
  \frac{2 C_a}{\dm} \hat\sigma,
\end{equation}
with the kinematic constraint
\begin{equation}
  -\log\left(\frac{M}{\sqrt{s}}\right) \le y 
  \le \log\left(\frac{M}{\sqrt{s}}\right)-\log\left(1+\frac{\dm}{M}\right)
\end{equation}

In the limit of small $\dm$, \ie neglecting the $\dm$ offset in the
PDF, one gets
\[
\frac{{\rm d}\sigma_a^{NLO}}{{\rm d}\dm}(y) \approx 
\frac{\alpha_s C_R}{\pi^2}\,\frac{1}{\dm}\, \sigma_a^{LO}(y),
\] 
which is exactly the result we have obtained in the soft limit up to
the geometric factor\footnote{This corresponds to the integration over
the geometrical region in $y$ and $\phi$ where the gluon is emitted in
one of the two jets.} $2 A_i(R)$ that we will need to reintroduce at
the end of the computation.

To simplify the discussion, we shall assume that the PDFs take the form
\begin{equation}\label{eq:PDF}
f_a(x) = N_a\,x^{-\lambda-1}\,(1-x)^{\beta_a}.
\end{equation}
The coefficients $N_a$ and $\beta_a$ will depend on the type of parton
and the scale $M$ but we shall assume a unique value for $\lambda$.

At this level, we could proceed by integrating over the rapidity
$y$. This is feasible analytically with the choice of PDF
\eqref{eq:PDF} but it leads to a Gauss hypergeometric function in
$\dd P^{(0)}_{i,f}/\dd\dm$ for which the Sudakov factor
\eqref{eq:sudakov} cannot be computed analytically. We will therefore
carry on with the computation at a fixed rapidity $y$.

Taking into account the emissions from both partons entering the
collisions, normalising to the $LO$ cross-section and reinserting the
geometric factor, one obtains
\begin{equation}\label{eq:isrfullbare}
\frac{{\rm d}P_i^{(0)}}{{\rm d}\dm}
 = \frac{\alpha_s C_a}{\pi^2}\,A_i(R)\,\frac{1}{\dm}
   \tau^{-\lambda-1}\left[
     \left(\frac{1-\kappa\tau e^  y }{1-\kappa e^  y }\right)^{\beta_a}+
     \left(\frac{1-\kappa\tau e^{-y}}{1-\kappa e^{-y}}\right)^{\beta_a}
   \right],
\end{equation}
where we have introduced $\kappa=M/\sqrt{s}$ and $\tau=1+\dm/M$, and
it is understood that the terms in the squared brackets are restricted
to the appropriate phase space \ie $\kappa\tau e^{\pm y}\le 1$,
respectively for each term.

In order to compute the Sudakov form factor analytically, we shall
make the approximation that the exponents $\beta_a$ are integers (in
practice, we shall use $\beta_q=3$ and $\beta_g=5$ for (anti-)quark and
gluon-jets respectively), and expand the small-$x$ power behaviour to
first order in $\dm$:
\begin{equation}
\tau^{-\lambda-1} \approx 1-(1+\lambda)\frac{\dm}{M}.
\end{equation}
Using
\begin{equation}
\frac{1-\kappa\tau e^{\pm y}}{1-\kappa e^{\pm y}}
 = 1 - \frac{\kappa e^{\pm y}}{1-\kappa e^{\pm y}}\,\frac{\dm}{M},
\end{equation}
the fact that $\beta_a$ is an integer allows one to see
\eqref{eq:isrfullbare} as a polynomial in $\dm$. 

We can then easily compute the Sudakov factor using the following
fundamental integrals:
\begin{eqnarray}
\int_{\dm}^{M/2}\frac{{\rm d}u}{\log(u/\Lambda_i)}\,u^{p-1} 
  & \overset{p=0}{=} & \log\left\lbrack \frac{\log(M/(2\Lambda_i))}{\log(\dm/\Lambda_i)} \right\rbrack, \\
  & \overset{p\neq 0}{=} &
  \Lambda_i^p\left\{{\rm Ei}\left[p\,\log(M/(2\Lambda_i))\right] - {\rm Ei}\left[p\,\log(\dm/\Lambda_i)\right]\right\},
\end{eqnarray}
where ${\rm Ei}(x)$ is the exponential integral.

With these simplifying assumptions, the final perturbative spectrum
for initial-state radiation can be written as
\begin{equation}\label{eq:isrfull}
  \frac{{\rm d}P_i}{{\rm d}\dm}
  = \frac{K_i}{\log(\dm/\Lambda_i)}\,\left(\sum_{k=0}^{\beta_a+1} \mu_k\,\frac{\dm^{k-1}}{M^k}\right)\,P_i(\Lambda_i,\dm)
\end{equation}
with the integrated probability
\begin{equation}\label{eq:isrPfull}
  P_i(\Lambda_i,\dm)
  = \left[\frac{\log(\dm/\Lambda_i)}{\log(M/(2\Lambda_i))}\right]^{K_i}
\,\exp\left[-K_i\sum_{k=1}^{\beta_a+1} \mu_k \, {\cal E}_k \left(\frac{\Lambda_i}{M}\right)^k\right],
\end{equation}
and the coefficients
\begin{eqnarray}
\mu_k & = & \nu_k-(1+\lambda)\nu_{k-1},\\
\nu_k & = & \frac{1}{2} C_{\beta_a}^k
  \left[
    \left(\frac{\kappa}{\kappa\!-\!e^{-y}}\right)^k\,\Theta\!\left(\frac{\kappa\!-\!e^{-y}}{\kappa}\!-\!\frac{\dm}{M}\right) +
    \left(\frac{\kappa}{\kappa\!-\!e^  y }\right)^k\,\Theta\!\left(\frac{\kappa\!-\!e^  y }{\kappa}\!-\!\frac{\dm}{M}\right)
  \right],\\
\nu_{-1} & = & \nu_{\beta_a+1} = 0,\\
{\cal{E}}_k & = & {\rm Ei}\left[k\,\log(M/(2\Lambda_i))\right] - {\rm Ei}\left[k\,\log(\dm/\Lambda_i)\right].\label{eq:suduptoM2}
\end{eqnarray}
Note that to obtain (\ref{eq:suduptoM2}), we have to integrate $\dm$
up to $M/2$, which is strictly valid only when $|y| \le
\log(\kappa/2)$, but similar results can easily be derived at forward
rapidities.

% comparison with Monte-Carlo
%%%%%%%%%%%%%%%%%%%%%%%%%%%%%%%%%%%%%%%%%%%%%%%%%%%%%%%%%%%%%%%%%%%%%%%%%%
\subsection{Comparison with Monte Carlo} \label{sec:pqcd_mc}

\begin{figure}
  \includegraphics[angle=270,width=\textwidth]{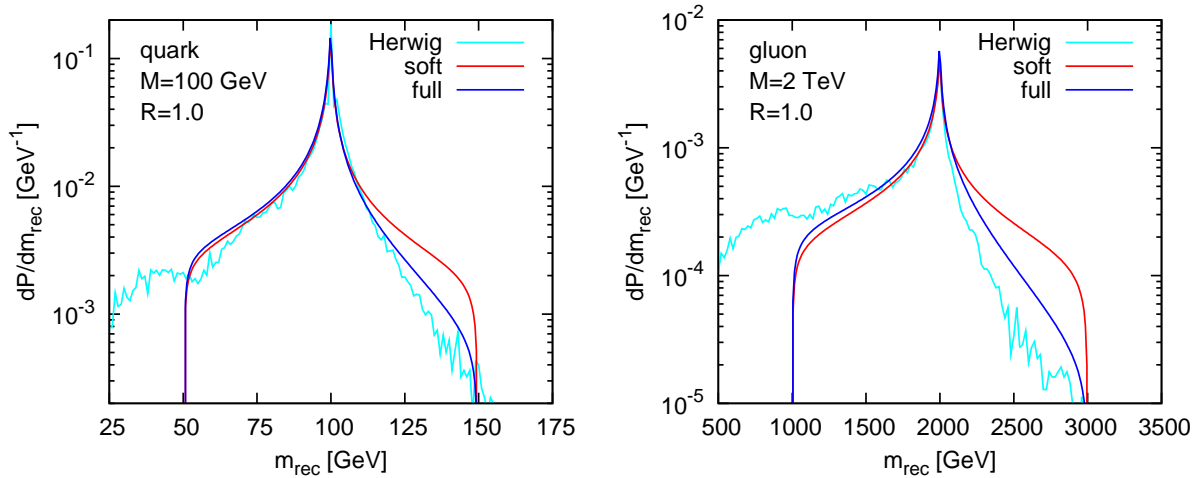}
  \caption{Reconstructed mass spectrum from our perturbative-QCD
  computation compared with Herwig at parton level (cyan curve). For
  the analytic computation, we show both the soft approximation (red
  curve) and the full case including PDF effects (blue curve). The
  left plot is for a $q\bar q$ dijet system at a nominal mass of 100
  GeV while the right plot is for the gluonic case at 2 TeV. In both
  cases, the anti-$k_t$ algorithm with a radius of 1 has been used
  for the clustering.}
  \label{fig:parton_hist}
\end{figure}

We want to conclude this Section by a comparison of the predictions we
obtain for the optimal radius from our analytical studies and from
Monte-Carlo approaches.
When dealing with the analytic result, we shall use the initial and
final-state results computed in the previous sections in the two
different situations for which we have analytic results: the soft
limit with a running-coupling approximation,
eqs. \eqref{eq:spectrum_pQCD_RC} and \eqref{eq:proba_pqcd_rc}, and the
``full'' spectrum where we also include the PDF effects in the
initial-state radiation, eq. \eqref{eq:isrfull}.

The total perturbative QCD spectrum is the convolution of the initial
and final-state pieces which, in practice, will be carried
numerically. We shall use $\Lambda_{\rm QCD}=200$ MeV and $N_f=5$ to
compute the running of the QCD coupling, and, for the PDF effects in
the initial-state-radiation spectrum, we shall adopt $\lambda=0.3$,
$\beta_q=3$ and $\beta_g=5$, and assume $y=0$, where we have checked
that most of the
massive objects in the Monte Carlo sample were produced. The quality measure is computed from the
spectrum using \eqref{eq:Qconstraints}.

\begin{figure}
  \includegraphics[angle=270,width=\textwidth]{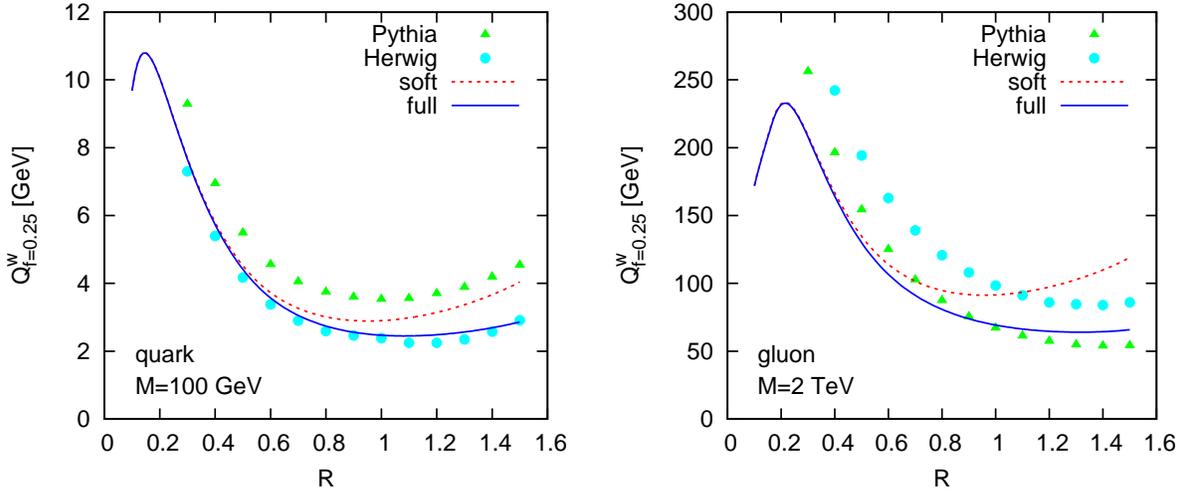}
  \caption{Quality measure as a function of $R$ for the parton-level
    spectrum. The (red) dashed curve corresponds to our analytic
    computation in the soft limit while the (blue) solid one includes
    PDF effects. The (green) triangles and (cyan) circles show the
    Monte-Carlo simulations for Pythia and Herwig respectively. As for
    Fig. \ref{fig:parton_hist}, the left plot is for a $q\bar q$ dijet
    system at a nominal mass of 100 GeV while the right plot is for
    the gluonic case with $M=2$ TeV.}
  \label{fig:parton_quality}
\end{figure}

We start our comparison directly with the spectrum for the
reconstructed dijet mass, as shown on Fig. \ref{fig:parton_hist} for
quark jets, $M=100$ GeV and gluon jets, $M=2$ TeV. Generally speaking,
the agreement between our analytic computation and Herwig is good,
even very good for the 100 GeV quark-jet case. Note that this
agreement is always better in the region of the peak, where the
soft-gluon emissions approximation we are working with is supposed to
hold. Similarly the small differences in the tail at small and large
$m_{{\rm rec}}$ are beyond the scope of our approximation. In both the
quark and gluon-jet cases, we see that the inclusion of PDF effects,
reducing the initial-state radiation, significantly improves the
description, especially in the case of large-mass gluon jets, where we
are sensitive to fast-falling gluon distributions at large $x$.

\begin{figure}
  \includegraphics[angle=270,width=\textwidth]{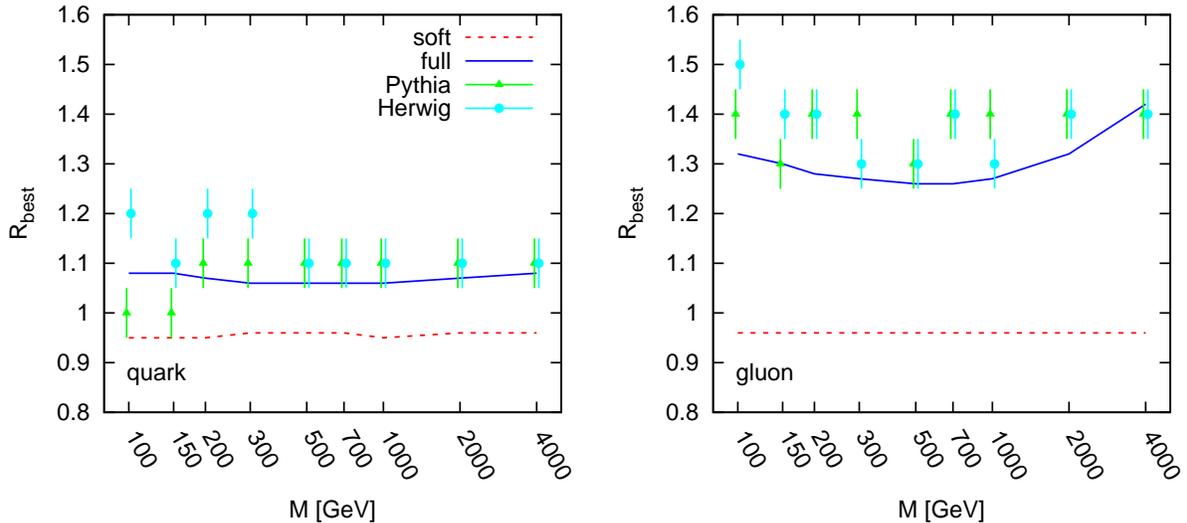}
  \caption{Best radius as a function of the mass, obtained from our
    analytic computation and compared to Monte-Carlo results. The
    analytic curves correspond to the soft limit (dotted red curve),
    and full computation (solid blue curve). These are compared to
    values of \Rbest extracted from parton-level Pythia (green
    triangles) and Herwig (cyan circles) simulations. The left plot
    shows the case of a resonance decaying into $q\bar q$ jets while
    the right plot has gluon jets.}
\label{fig:parton}
\end{figure}

Next, we turn to the computation of the quality measure, presented in
Fig. \ref{fig:parton_quality}. After properly
taking into account the PDF effects, our computation agrees in shape
with the Monte-Carlo results. For the normalisation, we are closer to
the Herwig simulation in the case of quark jets, while we better agree
with Pythia simulations for gluon jets. Since our analytic computation
is identical in both cases, it it somehow hard to find an explanation
for those differences.

Finally, let us concentrate on the values of \Rbest extracted from the
previous examples. This is shown on Fig. \ref{fig:parton} for quark
jets (left panel) and gluon jets (right panel). Despite the apparent
normalisation issue between Pythia and Herwig observed on
Fig. \ref{fig:parton_quality}, the fact that they agree on the shape
of the quality means that they yield very similar extracted optimal radii. If we turn
to the analytic computation, we first see that in the soft
approximation, the optimal radius is independent on the
process under consideration with $\Rbest\approx 0.96$ regardless of
the mass or the parton type\footnote{This is a bit smaller than the
  value around 1.1 expected from our discussion related to
  Fig. \ref{fig:coefs}. In practice, this is due to the fact that the
  quality measure is a complicated function of $A_{i,f}(R)$ involving
  \eg their derivatives which are different at $R=1.1$.}. The
reduction of the initial-state radiation due to the PDF effects has a
sizeable effect on the extracted optimal radius in such a way that, at
the end of the day, our analytic extraction of \Rbest in the ``full''
model reproduces very well the Monte-Carlo simulations with an error
$\lesssim$ 0.1.

%%%%%%%%%%%%%%%%%%%%%%%%%%%%%%%%%%%%%%%%%%%%%%%%%%%%%%%%%%%%%%%%%%%%%%%%%%
%% inclusion of the UE
%%%%%%%%%%%%%%%%%%%%%%%%%%%%%%%%%%%%%%%%%%%%%%%%%%%%%%%%%%%%%%%%%%%%%%%%%%
\section{Effects of the Underlying Event} \label{sec:UE}

% description of the effects
%%%%%%%%%%%%%%%%%%%%%%%%%%%%%%%%%%%%%%%%%%%%%%%%%%%%%%%%%%%%%%%%%%%%%%%%%%
\subsection{Description of the Underlying Event} \label{sec:full}

One of the most important effects when clustering jets at hadronic
colliders is the presence of the underlying event (UE) coming from
multiple interaction and beam remnants. This tends to produce a fairly
uniform soft radiation of a few GeV per unit area that contaminates the
jets and affects their momentum reconstruction.
In this Section, we therefore study the effects of this UE on the mass
spectrum we have computed from perturbative QCD in the previous Section.

If we work within the assumption that the UE is uniform in rapidity
and azimuth, the average $p_t$ contamination to each jet will be
${\cal A}_{\rm jet} \rho$ where ${\cal A}_{\rm jet}$ is the (active)
area of the jet \cite{areas} and $\rho$ the UE density per unit area.
For a given event, this contamination is affected by two kinds of
fluctuations: first, the fluctuations of the background itself which
go like $\sqrt{{\cal A}_{\rm jet}} \sigma$, with $\sigma$ the
background fluctuations per unit area.
Then, keeping in mind the possibility to discuss our results for
different jet algorithms than the anti-$k_t$ algorithm, the active
area of a jet is in general known to fluctuate, with an average value
proportional to $\pi R^2$ that increases with $p_t$ in a different way
for each jet algorithm.

For a given background density $\rho$, let us denote by ${\rm
d}P_i/{\rm d}p_{t,k}$ the probability distribution for a jet $k$ to
receive a contamination $p_{t,k}$ from the UE. For a jet of area
${\cal A}_{\rm jet}=\pi R^2\,a_k$, this distribution has an average of $\pi R^2\,a_k\rho$
and a dispersion $\sqrt{\pi R^2\,a_k}\sigma$. We therefore also have
to introduce ${\rm d}P_a/{\rm d}a_k$ the distribution, of average
$\bar{a}$ and dispersion $\sigma_a$ of the area divided by $\pi R^2$, of
the jet $k$.

If the two jets were perfect circles of radius $R$, one could easily
compute that the offset on the reconstructed mass would
be\footnote{The Bessel function mostly come from the fact that
  particles emitted far from the jet axis will contribute more to its
  energy.} $\dm = 2\pi R I_1(R) (\rho_1+\rho_2)$, up to corrections
proportional to $\dm^2/M$, where $I_1(x)$ is the modified Bessel
function of the first kind. For generic jets we shall then use
\begin{equation}\label{eq:dmfromUE}
\dm = \frac{2I_1(R)}{R} (p_{t,1}+p_{t,2}),
\end{equation}
where $p_{t,1}$ and $p_{t,2}$ are the transverse momenta of the
background contamination of our dijets.

On top of that, when we consider an event sample, the values of $\rho$
and $\sigma$ will fluctuate from one event to another. For the
event-by-event fluctuations of the background densities, we shall
assume a distribution ${\rm d}P_e/{\rm d}\rho$, of average
$\avg{\rho}$ and dispersion $\sigma_\rho$. For the intra-event
fluctuations (per unit area) $\sigma$ we shall simply assume that they
remain proportional\footnote{In terms of the toy model for the
  underlying event introduced in \cite{uepaper}, this corresponds to a
  fixed density of particles with a varying mean $p_t$.}  to $\rho$,
\ie, $\mu = \sigma/\rho = \avg{\sigma/\rho}$.

Given the event-by-event $\rho$ distribution ${\rm d}P_e/{\rm d}\rho$,
the intra-event fluctuations, ${\rm d}P_i/{\rm d}p_{t,k}$ and the area
distribution ${\rm d}P_a/{\rm d}a_k$, the UE spectrum can be inferred
from (\ref{eq:dmfromUE})
\begin{equation}\label{eq:exactuedistrib}
\frac{{\rm d}P_{\rm UE}}{{\rm d}\dm} = 
\int {\rm d}\rho \frac{{\rm d}P_e}{{\rm d}\rho}
\int {\rm d}a_1 {\rm d}a_2 \frac{{\rm d}P_a}{{\rm d}a_1} \frac{{\rm d}P_a}{{\rm d}a_2}\,
\int {\rm d}p_{t,1} {\rm d}p_{t,2} \frac{{\rm d}P_i}{{\rm d}p_{t,1}} \frac{{\rm d}P_i}{{\rm d}p_{t,2}}
\delta\left(\dm - \frac{2I_1(R)}{R} (p_{t,1}+p_{t,2})\right).
\end{equation}

Without knowing exactly that distribution, we can compute its average
and dispersion in terms of the properties of the fundamental
distributions:
\begin{eqnarray}
\avg{\dm}
   & = & \int {\rm d}\dm\,\dm\,\frac{{\rm d}P_{\rm UE}}{{\rm d}\dm}\label{eq:ueavg}\\
   & = & \left[\frac{2I_1(R)}{R}\right]\,2 \pi R^2 \bar{a}\,\avg{\rho},\nonumber\\
\sigma_{\dm}^2
   & = & \int {\rm d}\dm\,(\dm-\avg{\dm})^2 \,\frac{{\rm d}P_{\rm UE}}{{\rm d}\dm}\label{eq:uedisp}\\
   & = & \left[\frac{2I_1(R)}{R}\right]^2 \left[
           4 \bar{a}^2 (\pi R^2)^2 \sigma_\rho^2
           + 2 (\pi R^2)^2 \left(\avg{\rho}^2+\sigma_\rho^2\right) \sigma_a^2
           + 2 \bar{a} (\pi R^2)
           \left(\avg{\rho}^2+\sigma_\rho^2\right) \mu^2
         \right].\nonumber
\end{eqnarray}
The average is consistent with what one would naively expect, and the
dispersion is a combination of the three separate sources of
dispersion we are including: event-by-event fluctuations in $\rho$
(the first term), area fluctuations (the second term) and intra-event
fluctuations (the third term). As expected, the first two are
proportional to the area of the jet \ie to $\pi R^2$, while the
intra-event fluctuations are proportional to its square root.

So, instead of modelling the various distributions separately and
perform the integrations in \eqref{eq:exactuedistrib}, we shall
directly model the total UE spectrum and adjust its parameters to
reproduce the total average and dispersion. More specifically, we
shall assume
\begin{equation}\label{eq:spectue}
\frac{{\rm d}P_{UE}}{{\rm d}\dm}
  = \frac{1}{\Gamma(\alpha)} \frac{\dm^{\alpha-1}}{\dm_0^\alpha}e^{-\dm/\dm_0}.
\end{equation}
The parameters $\dm_0$ and $\alpha$ can be adjusted to reproduce the
average (\ref{eq:ueavg}) and dispersion (\ref{eq:uedisp}) of the UE
distribution. One finds
\begin{equation}
\alpha = \frac{\avg{\dm}^2}{\sigma_{\dm}^2}
\qquad\text{ and }\qquad 
\dm_0 = \frac{\sigma_{\dm}^2}{\avg{\dm}}.
\end{equation}
This choice has the advantage over a Gaussian distribution that it
remains positive.

% comparison with Monte-Carlo
%%%%%%%%%%%%%%%%%%%%%%%%%%%%%%%%%%%%%%%%%%%%%%%%%%%%%%%%%%%%%%%%%%%%%%%%%%
\subsection{Computation and comparison with Monte Carlo} \label{sec:full_mc}

Our strategy is rather straightforward: as we did when combining the
initial-state and final-state radiation spectra, we shall convolute
the perturbative QCD spectrum, discussed in Section \ref{sec:pqcd_mc},
with the UE spectrum \eqref{eq:spectue}. 
For each mass and parton-type, we wish to compare our results with the
Monte-Carlo studies \cite{optimisation}.

For the perturbative QCD part of the spectrum, we use the convolution
already discussed in Section \ref{sec:pqcd_mc} and consider the same
models: the pure soft approximation and the spectrum including PDF
effects in the initial-state. The result is itself convoluted with the
underlying-event distribution for which we still have to determine the
parameters appearing in eqs. (\ref{eq:ueavg}) and
(\ref{eq:uedisp}). The area properties $\bar{a}$ and $\sigma_a$ have
been calculated, together with their scaling violations in
\cite{areas}. The properties of the background are computed directly
from the Monte-Carlo event sample using FastJet \cite{fastjet} and the
method advocated in \cite{subtraction} and \cite{uepaper}: each event
is first clustered with the Cambridge/Aachen algorithm with $R=0.6$;
$\rho$ and $\sigma$ are then estimated from the jets that are within a
rapidity strip of 1 unit around the hardest dijet system, excluding
the two hardest jets in the event. The average properties
$\avg{\rho}$, $\mu$ and $\sigma_\rho$ can then be easily extracted
from the complete event sample. For completeness, we give a table of
the values we have obtained in Appendix \ref{app:ueprops}.

With that method, we are able to compute the mass spectrum, and thus the
quality measure, at a given mass and with a given parton type and UE
characteristics. We then extract \Rbest by searching the minimum of
the quality measure by steps of 0.01 in $R$.

\begin{figure}
  \includegraphics[angle=270,width=\textwidth]{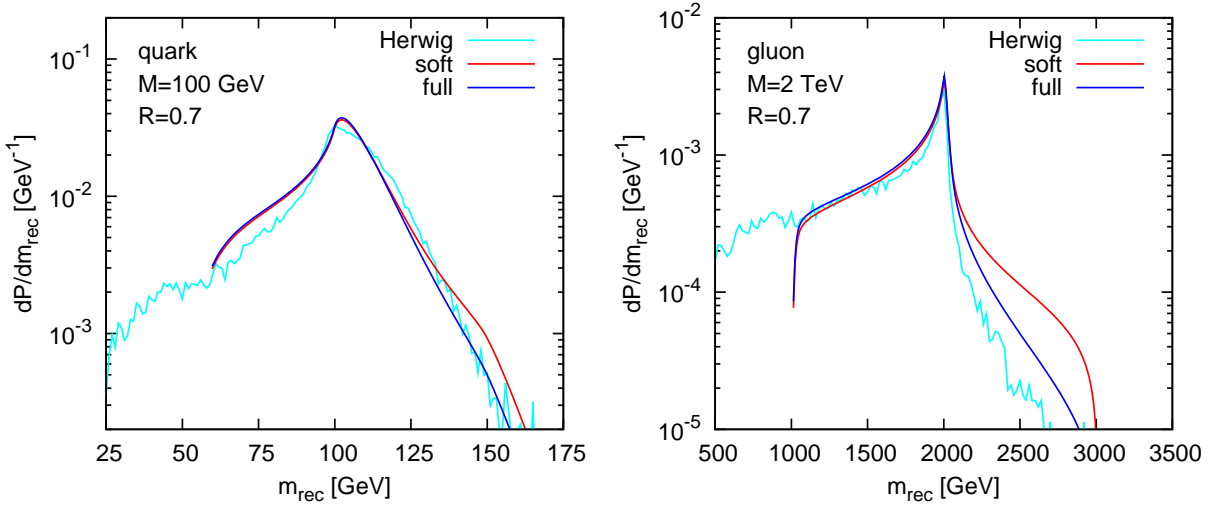}
  \caption{Reconstructed mass distribution for quark dijets at 100 GeV
    (left panel) and gluon dijets at 2 TeV (right panel). We compare
    the Monte-Carlo simulations from Herwig with our analytic
    computation. In both cases, we have used $R=0.7$ and the
    anti-$k_t$ algorithm.}
  \label{fig:ue_hist}
\end{figure}

\begin{figure}
  \includegraphics[angle=270,width=\textwidth]{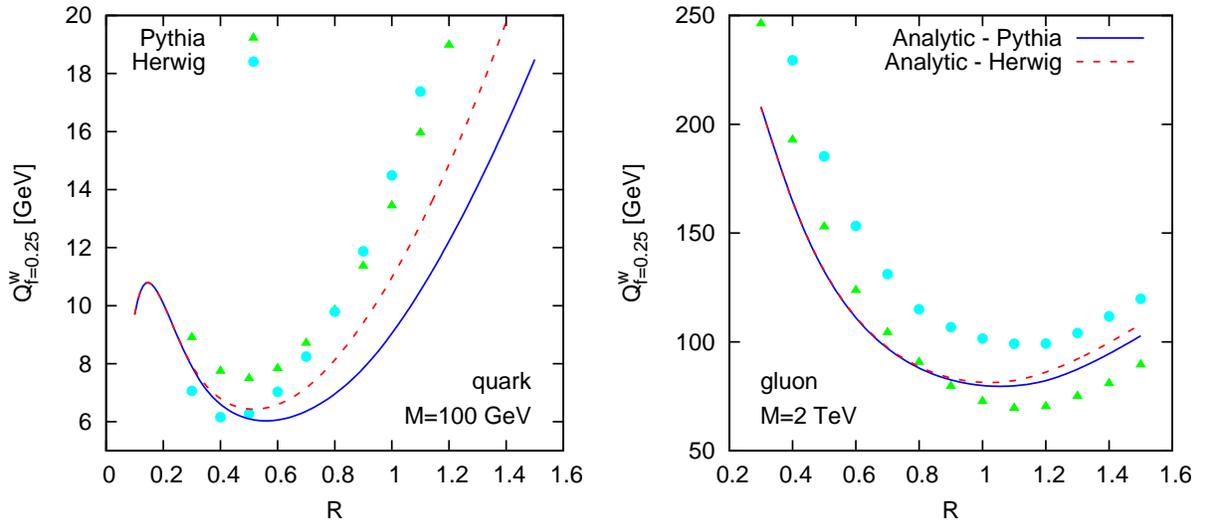}
  \caption{Quality measures as a function of $R$ for quark dijets at
    100 GeV (left panel) and gluon dijets at 2 TeV (right panel). The
    triangles and solid lines correspond to the Pythia simulations
    while the circles and dashed lines correspond to Herwig
    simulations.}
  \label{fig:ue_quality}
\end{figure}

\begin{figure}
  \includegraphics[angle=270,width=\textwidth]{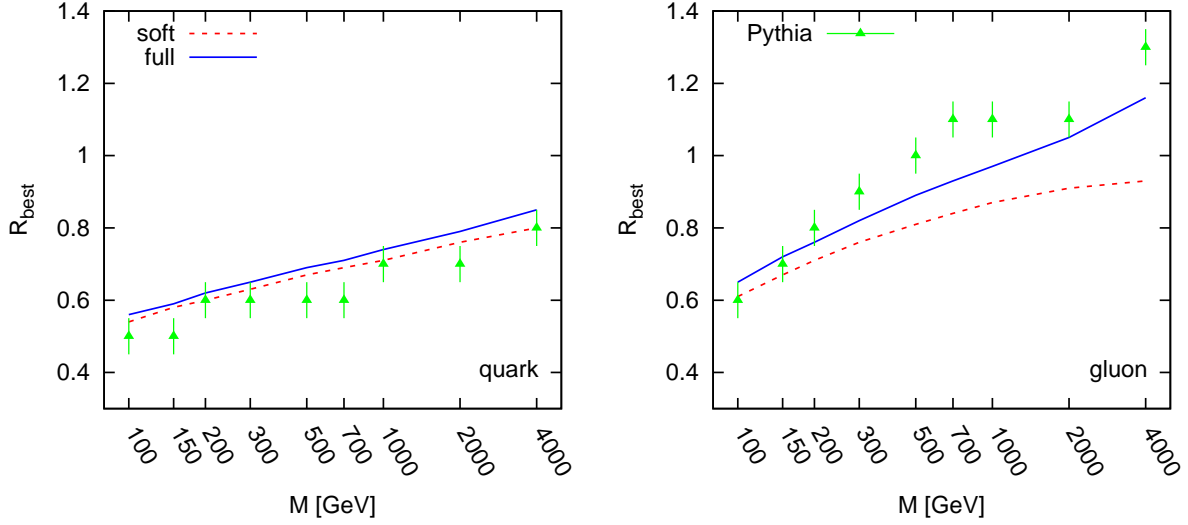}
  \caption{Best radius as a function of the mass when including the
    UE. Our analytic results are compared with Pythia Monte-Carlo
    simulations. The left (right) plot is for quark (gluon) jets.}
  \label{fig:ue_pythia}
\end{figure}

\begin{figure}
  \includegraphics[angle=270,width=\textwidth]{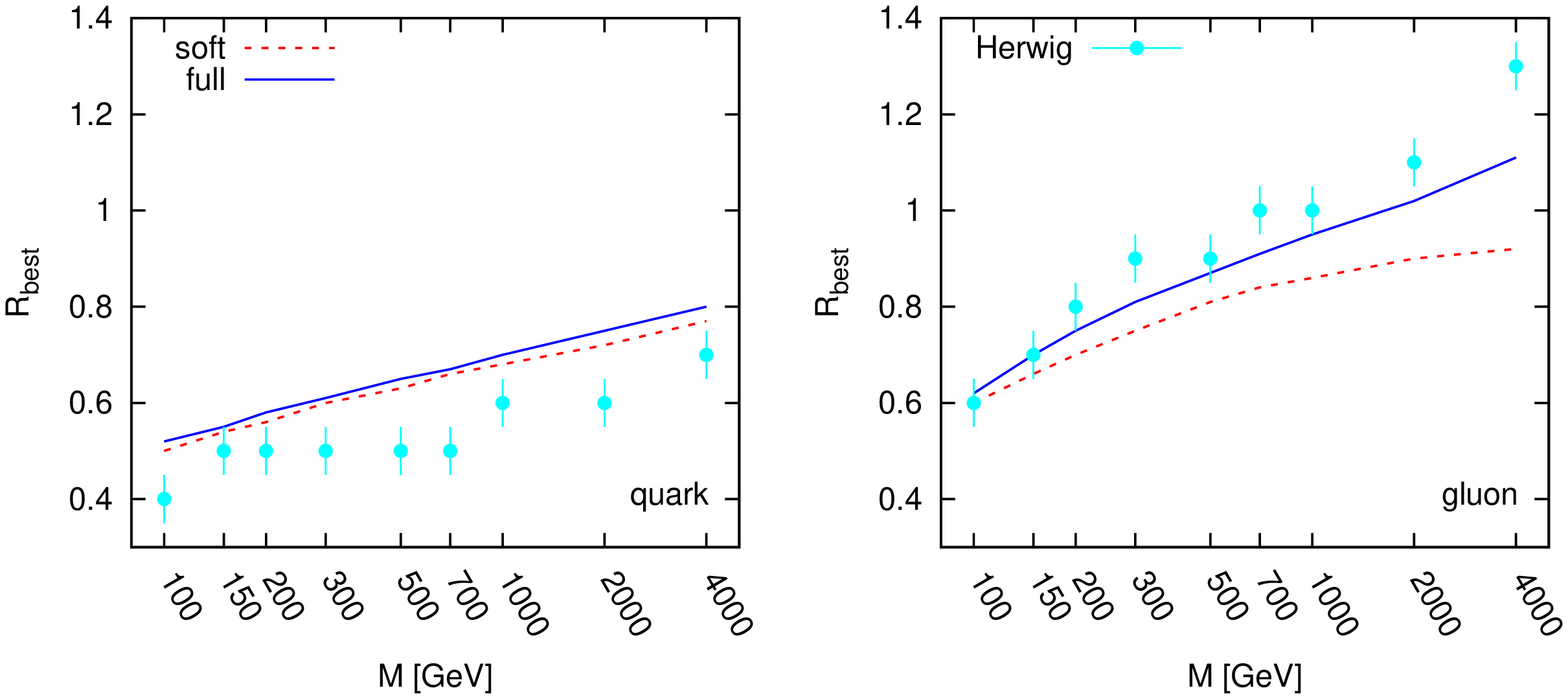}
  \caption{Same as Fig. \ref{fig:ue_pythia} but for Herwig simulations.}
  \label{fig:ue_herwig}
\end{figure}

As for the parton-level case, we shall compare our computations with
the Monte Carlo simulations from Pythia (tune DWT) and Herwig (with
default Jimmy tune) as the details of the Underlying Event can
differ. We shall again proceed in three steps: first consider the
histograms, then the quality measures and finally the optimal
radius. For simplicity, we shall focus on the anti-$k_t$ algorithm for
which $\bar{a}=1$ and $\sigma_a^2=0$.

Let us start with the histograms, Figure \ref{fig:ue_hist}, where we
illustrate our description of the spectrum for the quark dijets at
small mass and gluon dijets at large mass. In both cases, our analytic
histogram correctly reproduces the behaviour observed with Herwig. We
note however that our analytic computation slightly underestimating
the UE in the peak region in the case of quark jets at small nominal
mass\footnote{In the case of Pythia simulations, this underestimation,
  still observable in the case of quark jets at small mass $M$, is a bit
  more pronounced.} $M$. If we then move to the case of the
quality measure, Figure \ref{fig:ue_quality}, we see that, given the
differences we have already observed at the parton level (see Figure
\ref{fig:parton_quality}), the addition of the underlying event gives
a reasonable description of the Monte-Carlo results, especially the
shape of the quality at large $R$. Note that only the curves
corresponding to the model including the PDF effects are shown on
Figure \ref{fig:ue_quality} for clarity reasons.

Finally, the extracted optimal radius is presented on Figures
\ref{fig:ue_pythia} and \ref{fig:ue_herwig} for Pythia and Herwig
simulations respectively, as a function of the mass for both quark and
gluon jets.
Generally speaking we obtain a good description of the Monte-Carlo
results. We see once again that the inclusion of the PDF effects
significantly improves the description of the optimal radius for gluon
jets at large mass, as expected. We slightly underestimate the optimal
radius in the case of gluon jets at large mass, but we have to notice
that this is the regime on which it matters the least as in that
region the minimum in the quality measure is rather flat.

We thus see that our complete analytic model, the solid (blue) curve on
Figures \ref{fig:ue_pythia} and \ref{fig:ue_herwig}, corresponding to
our analytic pQCD computation including PDF effects,
eq. (\ref{eq:isrfull}), convoluted with our model of the Underlying
Event, eq. (\ref{eq:spectue}), gives a good extraction of the
optimal radius \Rbest.

% comparison for other algorithms
%%%%%%%%%%%%%%%%%%%%%%%%%%%%%%%%%%%%%%%%%%%%%%%%%%%%%%%%%%%%%%%%%%%%%%%%%%
\subsection{Computation for other algorithms} \label{sec:full_algs}

\begin{figure}
  \includegraphics[angle=270,width=\textwidth]{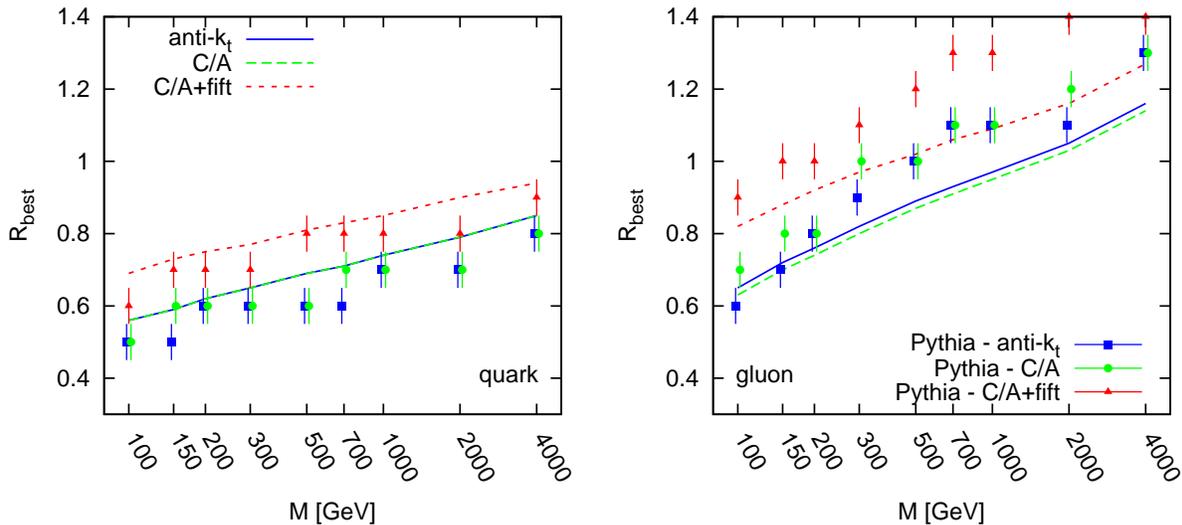}
  \caption{Best radius as a function of the mass for different
  algorithms. The lines are obtained using our analytic approach while
  the points correspond to Pythia simulations. The left plot shows the
  case of quark dijets and gluons are shown on the left panel. The
  solid (blue) lines together with the squares correspond to the
  anti-$k_t$ results, the dashed (green) lines and the circles show
  C/A results, and the dotted (red) lines and the triangles correspond
  to C/A with filtering.}
  \label{fig:algs_pythia}
\end{figure}

Now that we have checked the agreement between our analytic
computations and Monte-Carlo simulations for the simpler case of the
anti-$k_t$ algorithm, let us briefly discuss other possible jet
algorithms. In this section, we shall apply our analytic computations
to the case of the Cambridge/Aachen (C/A) as well as to the
Cambridge/Aachen algorithm supplemented by a filter (C/A+filt). While
the former usually leads to quality measurements close to what is
obtained in the anti-$k_t$ case, the latter produces narrower peaks
with an optimal radius slightly larger than the one obtained from C/A
without filtering. For simplicity, we shall follow what has been done
in \cite{optimisation} and use $R_{\rm filt}=R/2$ and $n_{\rm filt}=2$
for the parameters of the filter, \ie, recluster each jet with C/A and
a radius of half the original one and keep the two hardest subjets,
discarding the softer ones.

The analytic computation goes exactly as in the case of the anti-$k_t$
algorithm. The only thing that will change in our approach is the
inclusion of the area contribution in the UE distribution. The
expressions for the area are computed analytically, as in
\cite{areas}, up to order $\alpha_s$:
\begin{eqnarray}
\bar{a} & = &
  A_{0,{\rm JD}} +
  d_{{\rm JD}}\,\frac{C_a}{\beta_0\pi}\log\left[\frac{\alpha_s(Q_{0,{\rm JD}}^{(a)})}{\alpha_s(Rp_t)}\right],\\
\sigma_a^2 & = &
  \Sigma_{0,{\rm JD}}^2 +
  s_{{\rm JD}}^2\,\frac{C_a}{\beta_0\pi}\log\left[\frac{\alpha_s(Q_{0,{\rm JD}}^{(a)})}{\alpha_s(Rp_t)}\right],
\end{eqnarray}
where the coefficient $A_0$, $d$, $\Sigma_0$ and $s$ are computed in
perturbation theory and have been calculated\footnote{For the $k_t$, C/A
  and SISCone algorithms.} in \cite{areas}, while the scales $Q_0$ are
of non-perturbative origin and have been determined in \cite{areas}
from a fit to Herwig simulations. In our case, we shall use $p_t=M/2$
together with the following values for the coefficients\footnote{The
  coefficients for the filtered case have been computed with the same
  method as in \cite{areas} and we have simply assumed the same
  non-perturbative scales as for the C/A algorithm.} :
\begin{align}
A_{0,{\rm C/A}} = 0.814,\quad d_{\rm C/A}=0.083, &&
\Sigma_{0,{\rm C/A}}^2=0.0687,\quad s_{\rm C/A}^2=0.036,\\
A_{0,{\rm C/A+filt}} = 0.386,\quad d_{\rm C/A+filt}=0.019, &&
\Sigma_{0,{\rm C/A+filt}}^2=0.0046,\quad s_{\rm C/A+filt}^2=0.004,\\
Q_{0,{\rm C/A}}^{(q)}=Q_{0,{\rm C/A+filt}}^{(q)}=0.26\text{ GeV}, && Q_{0,{\rm C/A}}^{(g)}=Q_{0,{\rm  C/A+filt}}^{(g)}=0.53\text{ GeV}.
\end{align}

To avoid multiplying the figures, we just show, see Figure
\ref{fig:algs_pythia}, the final result of the extracted optimal
radius as a function of the mass for the three algorithms under
consideration, compared to Pythia simulations. We see that, in
agreement with the Monte Carlo, the differences between anti-$k_t$ and
Cambridge/Aachen are very small and thus our analytic approach
reproduces the expected behaviour correctly.  For the case of the C/A
algorithm with filtering, we obtain a good description in the case of
quark dijet systems but systematically underestimate the optimal
radius, by about 0.1-0.2, in the case of gluon jets. 
We believe that this is due to the fact that we should also introduce
some dependence on the algorithm at the level of the perturbative
spectrum. In our soft-gluon approximation all the algorithm will have
the same initial and final-state radiation spectra. However, with a
more complete treatment, \eg including non-global logarithms as in
\cite{mathieu}, differences will start appearing. Another effect comes
from the combination of the initial and final-state radiation spectra:
the simple convolution we have used is reasonable in the case of the
anti-$k_t$ algorithm where the shape of the jet is not affected by the
radiation, but more subtle phenomena will appear for more complex
algorithms, especially in the filtered case where, in some geometric
regions, the capture of ISR gluons can be affected by internal gluon
radiation in the jet.

All these effects are beyond the reach of the present analysis. As
expected from the previous discussion, discrepancies are more
pronounced in the case of gluon jets at large mass, where perturbative
radiation dominated the spectrum. Our computation nevertheless gives
reasonable results at the end of the day, given also that for gluon
dijets at larger mass, the minimum in the quality measure is less
pronounced than in other situations.

Finally, let us mention that the case of the $k_t$ algorithm \cite{kt}
is also very close to what we obtain for the anti-$k_t$ and C/A
algorithm, hence well reproduced by our calculation. For the SISCone
algorithm, differences in the perturbative spectrum also appear when
the radiated gluon is not soft, we therefore postpone that case for a
further study.

% comparison with previous approach
%%%%%%%%%%%%%%%%%%%%%%%%%%%%%%%%%%%%%%%%%%%%%%%%%%%%%%%%%%%%%%%%%%%%%%%%%%
\subsection{Comparison with earlier approaches} \label{sec:dms}

To some extent, the present calculation can be seen as an extended
version of what has been done in \cite{analytic}, where the authors
compute the average $p_t$ shift of a jet due to final-state radiation,
hadronisation and the UE. The approach we pursue in this paper is
however much more complete in the sense that we obtain the full
spectrum instead of just the average value\footnote{Also, we did not include
  hadronisation effects which are not important in the case we are
  concerned with, and we include initial-state radiation effects which
  are important to get the partonic picture right.}. As a consequence,
we can extract the width of the peak really as the dispersion in the
spectrum, while in \cite{analytic} the width is approximated by
\begin{equation}\label{eq:dmsdisp}
\avg{\delta p_t^2} \approx 
  \avg{\delta p_t}_{\rm pert}^2 +
  \avg{\delta p_t}_{\rm hadr}^2 +
  \avg{\delta p_t}_{\rm UE}^2
\end{equation}
\ie the total dispersion is approximated by the sum of each
independent average squared. The individual contributions are found to
be \cite{analytic}
\begin{eqnarray}
\avg{\delta p_t}_{\rm pert} & = & \frac{\alpha_s}{\pi}\log(R)L_ap_t,\nonumber\\
\avg{\delta p_t}_{\rm hadr} & = & \frac{-2 C_a}{R}{\cal A}(\mu_I),\\
\avg{\delta p_t}_{\rm UE}   & = & \Lambda_{\rm UE}\,RJ_1(R)\nonumber
\end{eqnarray}
with $2C_F{\cal A}(\mu_I)=0.5$ GeV for $\mu_I=2$ GeV and
\begin{eqnarray}
L_q & = & \left(2 \log(2)-\frac{3}{8}\right) C_F, \nonumber\\[-3mm]
&& \\[-3mm]
L_g & = & \left(2 \log(2)-\frac{43}{96}\right) C_A + \frac{7}{48}N_fT_R. \nonumber
\end{eqnarray}
The optimal radius is then obtained by minimising \eqref{eq:dmsdisp}
w.r.t. $R$. 

\begin{figure}
  \includegraphics[angle=270,width=\textwidth]{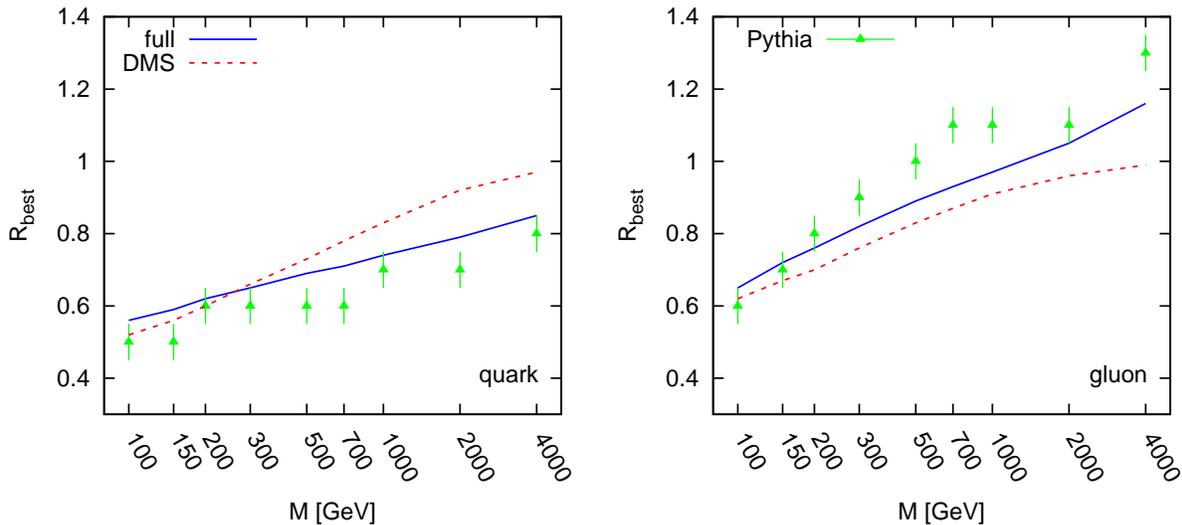}
  \caption{Best radius as a function of the mass for our approach
  compared to the previous work of \cite{analytic}. The solid (blue)
  line corresponds to our present computation in this paper while the
  dotted (red) line shows the result obtained in
  Ref. \cite{analytic}. The Pythia results, the (green) triangles, are
  shown for comparison. The left (right) panel shows the case of
  quark (gluon) jets.}
  \label{fig:dms}
\end{figure}

In practice, we have used $p_t = M/2$, $N_f=5$, $\Lambda_{\rm QCD}=200$ MeV and
computed the QCD coupling at the scale $Rp_t$. Instead of using a
fixed value for $\Lambda_{\rm UE}$ as in \cite{analytic}, we have used
the fact that it is related to the average background density,
$\avg{\rho}$, through (see Section \ref{sec:full})
\begin{equation}
\Lambda_{\rm UE} = 2 \pi \bar{a} \avg{\rho} \frac{I_1(R)}{J_1(R)},
\end{equation} 
and used the values extracted from the Monte Carlo event samples and
already used in Section \ref{sec:full} for $\avg{\rho}$.

The extracted value of \Rbest is shown on Fig. \ref{fig:dms} in the
case of Pythia simulations, for the anti-$k_t$ algorithm. The results
are compared with the Monte Carlo results and with our complete
analytic calculation.
We see that the simpler approach of \cite{analytic} manages to
extract the correct value for \Rbest at small mass but fails to
reproduce the observed behaviour at larger mass, and this for both
quark and gluon jets.
The reason for this failure is rather hard to pinpoint: it can come,
\eg, from the crudeness of the approximation (\ref{eq:dmsdisp}), the
lack of an initial-state radiation contribution and the corresponding
PDF effects, or the description of the underlying event in terms of
the average density $\avg{\rho}$ instead of its fluctuations
$\sigma_\rho$.

% subtracted case
%%%%%%%%%%%%%%%%%%%%%%%%%%%%%%%%%%%%%%%%%%%%%%%%%%%%%%%%%%%%%%%%%%%%%%%%%%
\section{Background subtraction} \label{sec:subtraction}

\subsection{Description} \label{sec:sub_descr}

The last point we wish to discuss is the situation where we perform
background subtraction using jet areas \cite{areas}, as advocated in
\cite{subtraction}. The idea behind UE subtraction is to get rid of
the shift due to the UE density contaminating the mass
reconstruction. Since subtraction is performed on each single event,
one hopes to get rid of the fluctuations of the background density
across the events, hence obtaining a narrower peak. In this respect,
all we should be left with is the background fluctuations inside an
event which have a linear dependence on $R$ rather than the quadratic
one computed in the unsubtracted case \eqref{eq:uedisp}. This is a
much smaller effect and one thus expect the quality measure to be
better (\ie smaller) than in the unsubtracted case as well as the
optimal radius to be larger.

In practice, one would thus naively convolute the perturbative
spectrum computed in Section \ref{sec:pqcd} with a subtracted UE
spectrum taking into account these intra-event fluctuations. Doing
that would actually lead to a quality much better than what is
observed in Monte-Carlo studies and to a systematic over-estimation of
\Rbest. 

The additional effect that is mandatory to take into account is the
fact that subtraction is performed using an estimated value for the
background density $\rho_{\rm est}$ which can differ from the exact
value of $\rho$ for that event. This results in some smearing effect
left over from the unsubtracted spectrum.
Because of this, the computation of the subtracted UE spectrum turns
out to be more involved than the corresponding unsubtracted case
computed in Section \ref{sec:full}, mostly because one also would have
to consider the distribution probability for $\rho_{\rm est}$. 

The properties of the estimated value of $\rho$ have actually recently
been discussed at length in \cite{uepaper}, where we learn that the
estimation of the background density using a median-based approach
suffers from two sources of bias: one of soft origin, related to the
fact that the median of a set of pure-UE jets may differ from its
average; and a contamination from radiation coming from the hard jets.
If we call $\rho$ and $\sigma$ the average UE density and its
fluctuations for one given event, and define $\delta=(\rho_{\rm
est}-\rho)/\rho$ as the relative error in the estimation of $\rho$,
the average and dispersion of $\delta$ are \cite{uepaper}
\begin{equation}
\avg{\delta} \simeq d_1\mu-d_2\mu^2
\quad\text{ and }\quad
\sigma_\delta^2 \simeq S_d^2\mu^2
\end{equation}
where, as in Section \ref{sec:full} we used $\mu=\sigma/\rho$, and the
coefficients are\footnote{For the hard contamination, we assume, as
we shall do in the next Section, that the ``bare'' jets have been
excluded from the range.}
\begin{eqnarray}
d_1 & = & R_\rho\sqrt{\frac{\pi c_J}{2}}\left(\frac{\avg{n_h}}{A_{\rm tot}}+c_J R_\rho^2\frac{\avg{n_h}^2}{2 A_{\rm tot}^2}\right),\\
d_2 & = & \frac{1}{4 c_J^2 R_\rho^2} 
\end{eqnarray}
and
\begin{equation}
S_d^2 = \frac{\pi}{2 A_{\rm tot}} + 2\pi c_J R_\rho^2\frac{\avg{n_h}}{A_{tot}^2} ,
\end{equation}
with
\begin{equation}\label{eq:nh}
\avg{n_h} = A_{\rm tot}\frac{C_a}{2\pi^2\beta_0}\log\left(\frac{\alpha_s({\rm min}(Q_0,\sqrt{c_J\sigma R_\rho}))}{\alpha_s(Q)}\right).
\end{equation}
In the previous set of equations, $c_J\simeq 2.04$ is a pure number
extracted in \cite{uepaper}, $R_\rho$ is the jet radius used to
estimate the background properties, $A_{\rm tot}$ is the total area of
the range we are using to estimate the background, $Q_0$ is a soft
cut-off that we fixed to 1 GeV, and $Q$ is the hard scale that we
naturally took equal to $M/2$.

These properties constrain the distribution of $\delta$, that we shall
denote by $\dd P_\delta/\dd \delta$, as a function of $\mu$. Since the
intra-event fluctuations will play a substantial role in this
UE-subtracted situation, we shall also consider a distribution for
them. For simplicity, we shall parametrise the distribution in terms
of $\mu$ rather than in terms of $\sigma$ as background estimations
are naturally expressed in that variable. We therefore introduce $\dd
P_\mu/\dd \mu$, of average $\avg{\mu}$ and dispersion
$\sigma_\mu^2$. In the course of the computation below, we shall
encounter higher cumulants for this distribution. To keep things
simple, we shall work with the Gaussian
assumption\footnote{Alternatively, one can use the moments obtained
  from a Beta distribution which would enforce the positivity of
  $\mu$. We have checked that this choice does not affect the
  resulting \Rbest.} \ie $\avg{\mu^3} =
\avg{\mu}(\avg{\mu}^2+3\sigma_\mu^2)$,
$\avg{\mu^4}=\avg{\mu}^4+6\avg{\mu}^2\sigma_\mu^2+3\sigma_\mu^4$. The
values of $\avg{\mu}$ and $\sigma_\mu^2$ can again be extracted from
the Monte-Carlo samples.

With this at hand, we proceed, as in Section \ref{sec:full}, by
computing the mass shift for a given configuration and deducing its
average and dispersion. For given values of $\rho$, $\mu$, $\rho_{\rm
est}$, jet areas $a_k$ and jet UE contaminations $p_{t,k}$, the offset
in the reconstructed mass after subtraction is
\begin{equation}
\dm = \left[\frac{2 I_1(R)}{R}\right] \left[p_{t,1}+p_{t,2}-\pi R^2 (a_1+a_2)(1+\delta)\rho\right],
\end{equation}
where the last term is the subtracted piece.

After a slightly lengthy computation we reach\footnote{The computation
  proceeds as in Section \ref{sec:full}. The only additional
  approximation is the replacement of $\sigma$ by
  $\avg{\mu}\avg{\rho}$ in $\avg{n_h}$ (eq. (\ref{eq:nh})), which only
  introduces subleading corrections.}
\begin{eqnarray}
\avg{\dm}
   & = & \left[\frac{2I_1(R)}{R}\right]\,2 \pi R^2 \bar{a}\,\avg{\rho}
         \left\lbrack -d_1\avg{\mu}+d_2(\avg{\mu}^2+\sigma_\mu^2)\right\rbrack,\label{eq:ueavgsub}\\
\sigma_{\dm}^2
   & = & \left[\frac{2I_1(R)}{R}\right]^2 \left(
\left\{\left[2\pi R^2 \bar{a}+2(\pi R^2)^2\sigma_a^2(d_1^2+S_d^2)+4(\pi R^2 \bar{a})^2 S_d^2\right](\avg{\rho}^2+\sigma_\rho^2)\right.\right.\nonumber\\[-5mm]
   &   & \phantom{\left[\frac{2I_1(R)}{R}\right]^2 ()}\left.+4(\pi R^2\bar{a})^2d_1^2\sigma_\rho^2\right\}(\avg{\mu}^2+\sigma_\mu^2)\nonumber\\[-5mm]
   &   & \phantom{\left[\frac{2I_1(R)}{R}\right]^2} \left. + \left[2(\pi R^2)^2\sigma_a^2(\avg{\rho}^2+\sigma_\rho^2)+4 (\pi R^2 \bar{a})^2\sigma_\rho^2\right]\left(d_2^2\avg{\mu^4}-2d_1d_2\avg{\mu^3}\right)\right.\nonumber\\[-5mm]
   &   & \phantom{\left[\frac{2I_1(R)}{R}\right]^2} \left. + \, 4 (\pi R^2 \bar{a})^2 \avg{\rho}^2\left[2 d_2^2\sigma_\mu^2+(2d_2\avg{\mu}-d_1)^2\right]\sigma_\mu^2\right).\label{eq:uedispsub}
\end{eqnarray}

Once we have these expressions, the last step is to parametrise the
subtracted UE spectrum. Since subtraction can produce negative values
for \dm, we cannot use \eqref{eq:spectue}. The technique we shall
adopt is to keep the same distribution using the unsubtracted average
together with the dispersion we have just computed, and then to shift
the complete distribution in order to reproduce the subtracted average
\dm. In practice, this gives
\begin{equation}\label{eq:spectuesub}
\frac{{\rm d}P_{UE,{\rm sub}}}{{\rm d}\dm}
  = \frac{1}{\Gamma(\alpha)} \frac{(\dm+\dm_1)^{\alpha-1}}{\dm_0^\alpha}e^{-(\dm+\dm_1)/\dm_0}.
\end{equation}
where the parameters $\dm_0$, $\dm_1$ and $\alpha$ are given by
\begin{equation}
\alpha = \frac{\avg{\dm}_{\rm unsub}^2}{\sigma_{\dm}^2},
\qquad
\dm_0 = \frac{\sigma_{\dm}^2}{\avg{\dm}_{\rm unsub}}
\qquad\text{ and }\qquad 
\dm_1 = \avg{\dm}_{\rm unsub}-\avg{\dm}.
\end{equation}
with $\avg{\dm}$ and $\sigma_{\dm}^2$ given by eqs. \eqref{eq:ueavgsub}
and \eqref{eq:uedispsub}, and $\avg{\dm}_{\rm unsub}$ by
eq. \eqref{eq:ueavg}.

Finally, it is interesting to notice that in the case of a perfect
estimation of the background density $\rho_{\rm est}=\rho$, \ie
$d_1=d_2=S_d=0$, one recovers
\begin{eqnarray}
\avg{\dm} & = & 0,\\
\sigma_{\dm}^2 & = & \left[\frac{2I_1(R)}{R}\right]^2 \, 2\pi R^2 \bar{a}(\avg{\rho}^2+\sigma_\rho^2)(\avg{\mu}^2+\sigma_\mu^2)
\end{eqnarray}
\ie no average shift and a (squared) dispersion that is proportional
to the area (linearly) and to $\sigma^2$.

\subsection{Comparison with Monte-Carlo simulations}\label{sec:sub_mc}

\begin{figure}
  \includegraphics[angle=270,width=\textwidth]{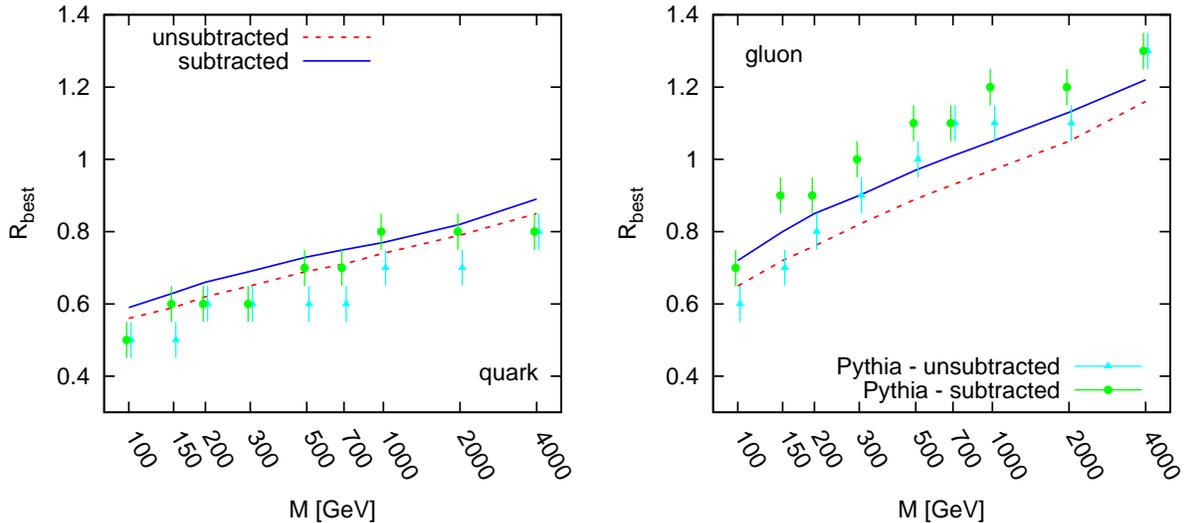}
  \caption{Best radius as a function of the mass for both the
    UE-unsubtracted and the UE-subtracted cases.  The solid (blue)
    line corresponds to our subtracted prediction and has to be
    compared to the Pythia Monte-Carlo simulation shown as (green)
    circles. For reference, we also show as a (red) dotted line the
    unsubtracted predictions, together with the Pythia results shown
    as the (cyan) triangles. The left (right) panel shows the case of
    quark (gluon) jets.}
  \label{fig:subtract}
\end{figure}

As for the previous cases we have analysed, we shall perform a
comparison of our analytic predictions with Monte Carlo simulations
for the UE-subtracted case. The method goes exactly as before, so we
just highlight the main steps and present our results for the optimal
radius \Rbest and stay with the anti-$k_t$ algorithm.

For the analytic part of the computation, the perturbative QCD
spectrum is convoluted with the subtracted UE spectrum calculated in
the previous Section. Concerning the parameters of the latter, all
background properties, $\avg{\rho}$, $\sigma_\rho^2$, $\avg{\mu}$ and
$\sigma_\mu^2$ are extracted directly from the Monte-Carlo data sample
and the estimation of the background density was performed using the
Cambridge/Aachen algorithm with a radius of 0.6. We have kept all the
jets within one unit of rapidity around the hardest dijet, excluding
the two hardest jets in the event, giving a total area $A_{\rm
  tot}=4\pi$. It is interesting to notice that when background
subtraction is performed, the final results will slightly depend on
the details of how $\rho$ is estimated, through the coefficients
$d_1$, $d_2$ and $S_d$ in our analytic approach.
Extending the range, \ie increasing its area $A_{\rm tot}$, would
decrease the bias in the estimation of $\rho$, but in that case, there
would be a risk that the background would not be uniform over the
range (\eg vary with rapidity).
Following
\cite{uepaper}, we could therefore also try to optimise the choice of
range and jet definition used
for estimating $\rho$.

The computed value of \Rbest is shown on Fig. \ref{fig:subtract} in
the case of Pythia simulations. For an easier comparison, we show both
the subtracted and unsubtracted results. For the analytic calculation,
we observe a value of \Rbest larger in the subtracted case by about
0.04 in the case of quark-jets and by about 0.08 in the case of
gluon-jets. All these values are in good agreement with what is
observed in the Monte-Carlo simulations.

%%%%%%%%%%%%%%%%%%%%%%%%%%%%%%%%%%%%%%%%%%%%%%%%%%%%%%%%%%%%%%%%%%%%%%%%%%
%% conclusions and perspectives
%%%%%%%%%%%%%%%%%%%%%%%%%%%%%%%%%%%%%%%%%%%%%%%%%%%%%%%%%%%%%%%%%%%%%%%%%%
\section{Conclusions and perspectives} \label{sec:ccl}

Throughout this paper we have investigated analytically the
optimisation of the radius $R$ one uses with a jet algorithm to
reconstruct dijet resonances.
We have already learnt \cite{optimisation} that carefully choosing the
jet radius can lead to significant improvement of the reconstructed
signal and we find it valuable to supplement previous Monte-Carlo studies
with an analytic understanding of these dijet reconstructions.

Our approach has been to calculate the reconstructed dijet mass
spectrum. This was done by convoluting fundamental pieces: the
perturbative QCD radiation in the initial and final state, and the
contamination due to the Underlying Event. 
For the perturbative part, we have worked in the soft-gluon emission
approximation, a choice motivated by the fact that we are mostly
interested in the behaviour in the vicinity of the mass peak. In the
case of the initial-state radiation spectrum, we have also included
PDF effects that can have a non-negligible impact, especially in the
case of gluon jets and at large dijet mass.
For the UE description, we have modelled it in such a way
as to reproduce the observed properties of the background.
From the reconstructed mass spectrum, we can compute a quality measure
and extract an optimal radius, following the method outlined in
\cite{optimisation}.

Our analytic results have been considered and compared with
Monte-Carlo simulations in various cases. We have started with
parton-level events, where we can test our analytic computation of the
perturbative spectrum; then we have focused on full hadronic events
including the Underlying Event; and, finally, we have considered the
situation with a jet-area-based background subtraction
\cite{subtraction}. We have mostly considered the case of the
anti-$k_t$ algorithm but have also considered other options.

Generally speaking, we can say that we obtain a good description of
the Monte-Carlo results, especially when including the PDF effects in
initial-state radiation. The optimal radius \Rbest computed
analytically is in most cases less than 0.1 away from the Monte-Carlo
expectation. The comparison at a more basic level, the quality measure
or even the mass spectrum, also shows a good agreement.

Most of the deviations between our analytic results and Monte
Carlo simulations can probably be traced back to what we observe at
the partonic level. Though our analytic computation reproduces nicely
the Monte-Carlo simulations, we have observed some deviations in
Section \ref{sec:pqcd_mc}. It is interesting to notice that the
differences between our description
and the Monte-Carlo results is usually of the same order as the
differences obtained between Pythia and Herwig.
These differences propagate to the situation where the UE is included.

The one case where our analytic results deviate a bit more from the
Monte-Carlo simulations is the case of the Cambridge/Aachen algorithm
supplemented by a filter (see Section \ref{sec:full_algs}), for events
including the UE, while our description remains good for the
anti-$k_t$ and the Cambridge/Aachen algorithm without any filter. We
think that a better description goes beyond the reach of the present
analysis, \eg already at the partonic level, a simple convolution
between the initial and final-state spectra may be too simplistic when
filtering is applied.

Let us also notice that our description of the UE allows us to treat
successfully both the case where the UE is subtracted and the case
where it is not. Our approach includes all the effects that contribute
to the dispersion of the reconstructed mass: mostly background
fluctuations from one event to another, intra-event fluctuations and
area fluctuations. It is interesting to notice that in the case of
subtracted UE, it was also mandatory to take into account the fact
that the estimated background can differ from its ``true'' value. This
leads to a significant additional contribution to the
dispersion. Using the analytic computations from \cite{uepaper} led to
good results for the extracted \Rbest.

If one wishes to improve the present description, additional effects
have to be included in the analytic computation: relaxing the
soft-gluon-emission approximation, obtaining sub-leading logarithmic
corrections, performing a better resummation of the soft virtual
corrections than a simple Sudakov factor, \eg using non-global
logarithms, and including hadronisation corrections. However, we have
seen that the description we can achieve with our simple approach is
already satisfactory without having to introduce these corrections. It
is nevertheless important to remember their existence as they might
become relevant in more specific situations, like the case of
filtering where we have observed that we start to reach the limits of
our simple approach. Some effects, like subleading logarithmic
corrections, may even be simpler to compute analytically than to
implement in a Monte Carlo.

Another point that we have not investigated and that might be
interesting in a future study is the case of the rapidity dependence
of \Rbest. As we have seen in Section \ref{sec:pdf}, PDF effects
introduce a dependence on the rapidity at which the heavy object is
produced \cite{trimming} and it might be interesting to see how \Rbest
varies with that rapidity.

Also, all the computations we have performed were for a complete
event-set. A next major step would be to optimise the radius on an
event-by-event basis. This is much more complicated as the previous
approach would then only give access to an average dispersion for a
single event and the additional dispersion arising when
combining the whole set of events would be much harder to handle. We
therefore also leave that step to future studies.

The question of knowing which jet algorithm should be preferred has
already triggered many discussions both on the theoretical and on the
experimental side, and more complex jet definitions introducing more
parameters have often been proposed to improve jet reconstruction. The
present work, together with \cite{mathieu}, can be seen as opening a new direction of improvement:
analytically optimising the parameters in the jet definitions. Considering a single
parameter, the jet radius, for a simple process, dijet
reconstructions, is a first step in that direction.
Gaining analytic control on the parameters entering jet definitions
can be extremely valuable. This is especially true when new parameters
are introduced: being able to predict them can avoid often costly
determinations based on Monte-Carlo simulations or rough empirical choices. The
application of the techniques developed in this paper to other
parameters and more general processes would therefore be very
interesting.

%%%%%%%%%%%%%%%%%%%%%%%%%%%%%%%%%%%%%%%%%%%%%%%%%%%%%%%%%%%%%%%%%%%%%%%%%%
%% conclusions and perspectives
%%%%%%%%%%%%%%%%%%%%%%%%%%%%%%%%%%%%%%%%%%%%%%%%%%%%%%%%%%%%%%%%%%%%%%%%%%
\subsection*{Acknowledgements}

I am extremely grateful to Gavin Salam for many useful suggestions and
comments at every step of this work. 
I also thank Lorenzo Magnea for a careful reading of the manuscript.
Finally, I thank the Brookhaven National Laboratory for support when
this work was initiated and the LPTHE (University of Paris 6) where
parts of this work has been done.

%%%%%%%%%%%%%%%%%%%%%%%%%%%%%%%%%%%%%%%%%%%%%%%%%%%%%%%%%%%%%%%%%%%%%%%%%%
%% appendices
%%%%%%%%%%%%%%%%%%%%%%%%%%%%%%%%%%%%%%%%%%%%%%%%%%%%%%%%%%%%%%%%%%%%%%%%%%
\begin{appendix}

%% extracted background properties
%%%%%%%%%%%%%%%%%%%%%%%%%%%%%%%%%%%%%%%%%%%%%%%%%%%%%%%%%%%%%%%%%%%%%%%%%%
\section{Extracted background properties}\label{app:ueprops}

\begin{table}
\begin{center}
\begin{tabular}{|r||ccc|ccc||ccc|ccc||}
\hline
 & \multicolumn{6}{c||}{Pythia}  & \multicolumn{6}{c||}{Herwig} \\
\cline{2-13}
 & \multicolumn{3}{c|}{quarks} & \multicolumn{3}{c||}{gluons}  & \multicolumn{3}{c|}{quarks} & \multicolumn{3}{c||}{gluons} \\
\cline{2-13}
mass & $\avg{\rho}$ & $\sigma_\rho$ & $\mu$ & $\avg{\rho}$ & $\sigma_\rho$ & $\mu$ & $\avg{\rho}$ & $\sigma_\rho$ & $\mu$ & $\avg{\rho}$ & $\sigma_\rho$ & $\mu$\\
\hline
 100 & 2.23 & 1.95 & 0.70 & 3.00 & 2.14 & 0.68 & 3.08 & 2.21 & 0.46 & 3.78 & 2.34 & 0.45 \\
 150 & 2.35 & 2.03 & 0.70 & 3.19 & 2.26 & 0.67 & 3.21 & 2.25 & 0.45 & 3.97 & 2.40 & 0.45 \\
 200 & 2.42 & 2.07 & 0.70 & 3.33 & 2.36 & 0.67 & 3.31 & 2.30 & 0.45 & 4.11 & 2.45 & 0.45 \\
 300 & 2.53 & 2.16 & 0.70 & 3.51 & 2.49 & 0.67 & 3.44 & 2.35 & 0.45 & 4.37 & 2.55 & 0.45 \\
 500 & 2.65 & 2.25 & 0.70 & 3.71 & 2.65 & 0.67 & 3.63 & 2.44 & 0.45 & 4.70 & 2.68 & 0.45 \\
 700 & 2.71 & 2.29 & 0.70 & 3.83 & 2.78 & 0.67 & 3.74 & 2.48 & 0.45 & 4.93 & 2.82 & 0.45 \\
1000 & 2.77 & 2.36 & 0.70 & 3.95 & 2.95 & 0.67 & 3.83 & 2.52 & 0.45 & 5.21 & 3.00 & 0.46 \\
2000 & 2.78 & 2.38 & 0.70 & 4.05 & 3.33 & 0.67 & 3.94 & 2.60 & 0.46 & 5.75 & 3.58 & 0.47 \\
4000 & 2.54 & 2.26 & 0.71 & 3.97 & 3.86 & 0.68 & 3.86 & 2.68 & 0.48 & 6.36 & 4.52 & 0.48 \\
\hline
\end{tabular}
\end{center}
\caption{Background properties needed to parametrise the UE
distribution. Values are given for quark and gluon jets as well as for
Pythia and Herwig+Jimmy simulations. All dimensionful numbers are expressed
in GeV.}\label{tab:ueprops}
\end{table}

For the sake of completeness, we briefly mention in this appendix the
values of the background properties that are necessary to fix the UE
spectrum. These will differ between quark and gluon jets and
between Pythia 6.4 (tune DWT) and Herwig 6.51 (with default Jimmy tune). All the values used in our analytic
computation are summarised in Table \ref{tab:ueprops}.

\end{appendix}

%%%%%%%%%%%%%%%%%%%%%%%%%%%%%%%%%%%%%%%%%%%%%%%%%%%%%%%%%%%%%%%%%%%%%%%%%%
%% bibliography
%%%%%%%%%%%%%%%%%%%%%%%%%%%%%%%%%%%%%%%%%%%%%%%%%%%%%%%%%%%%%%%%%%%%%%%%%%

\end{document}